\documentclass[fleqn,usenatbib]{mnras}

\usepackage{newtxtext,newtxmath}

\usepackage[T1]{fontenc}
\usepackage{ae,aecompl}
\usepackage{wasysym}

\usepackage{algpseudocode}
\usepackage{graphicx}	
\usepackage{amsmath}	
\usepackage{amssymb}	
\usepackage{multirow}
\usepackage[flushleft]{threeparttable}
\usepackage{multicol}
\usepackage{booktabs}
\DeclareMathOperator*{\argmin}{arg\,min}

\makeatletter
\newlength{\abovecaptionskip}%
\makeatother

\title[Subgiant properties with deep learning]{Asteroseismic Inference of Subgiant Evolutionary Parameters with Deep Learning}

\author[Hon et al.]{Marc Hon$^{1}$\thanks{E-mail: m.hon@unsw.edu.au},
Earl P. Bellinger$^{2,1,3}$,
Saskia Hekker$^{3,2}$,
Dennis Stello$^{1,2,4}$,
\newauthor
and James S. Kuszlewicz$^{3,2}$
\newauthor
\\
$^{1}$School of Physics, The University of New South Wales, Sydney NSW 2052, Australia\\
$^{2}$Stellar Astrophysics Centre, Department of Physics and Astronomy, Aarhus University, Ny Munkegade 120, DK-8000 Aarhus C, Denmark\\
$^{3}$Max-Planck-Institut f{\"u}r Sonnensystemforschung, Justus-von-Liebig-Weg 3, 37077 G{\"o}ttingen, Germany\\
$^{4}$Sydney Institute for Astronomy (SIfA), School of Physics, University of Sydney, NSW 2006, Australia
}

\date{Accepted XXX. Received YYY; in original form ZZZ}

\pubyear{2020}
\begin{document}
\label{firstpage}
\pagerange{\pageref{firstpage}--\pageref{lastpage}}
\maketitle

\begin{abstract}
With the observations of an unprecedented number of oscillating subgiant stars expected from NASA's TESS mission, the asteroseismic characterization of subgiant stars will be a vital task for stellar population studies and for testing our theories of stellar evolution. To determine the fundamental properties of a large sample of subgiant stars efficiently, we developed a deep learning method that
estimates distributions of fundamental parameters like age and mass over a wide range of input physics by learning from a grid of stellar models varied in eight physical parameters. 
We applied our method to four \textit{Kepler} subgiant stars and compare our results with previously determined estimates. Our results show good agreement with previous estimates for three of them (KIC 11026764, KIC 10920273, KIC 11395018). With the ability to explore a vast range of stellar parameters, we 
determine that the remaining star, KIC 10005473, is likely to have an age 1~Gyr younger than its previously determined estimate. Our method also estimates the efficiency of overshooting, undershooting, and microscopic diffusion processes, from which we determined that the parameters governing such processes are generally poorly-constrained in subgiant models. 
We further demonstrate our method's utility for ensemble asteroseismology by characterizing a sample of 30 \textit{Kepler} subgiant stars, where we find a majority of our age, mass, and radius estimates agree within uncertainties from more computationally expensive grid-based modelling techniques.
\end{abstract}




\section{Introduction}

Asteroseismology of solar-like oscillations is a powerful approach to measure ages of individual field stars. By probing the stellar interior, asteroseismic measurements can reveal structural changes that are indicators of stellar evolution. This is especially the case for subgiant stars that have begun to show mixed modes in their oscillation spectra. These modes arise from the coupling of acoustic waves that propagate in the stellar envelope with gravity ($g$-) waves that propagate near the core \citep{Osaki_1975}, and result in perturbations to the near-uniform frequency spacing of acoustic ($p$-) modes (\textit{avoided crossings}, \citealt{Aizenman_1977}). As the interiors of subgiants evolve over relatively short timescales, the mixed mode behaviour of the star's oscillation spectrum also changes rapidly (e.g., \citealt{JCD_1995}). Hence, detailed measurements of subgiant mixed modes not only provide valuable diagnostics of the stellar interior (e.g., \citealt{Deheuvels_2011,Benomar_2012, Benomar_2014}), but also yield precise stellar age estimates (e.g., \citealt{Deheuvels_2014, Metcalfe_2014, Li_2017}).



\begin{figure*}
    \centering
	\includegraphics[width=2.1\columnwidth]{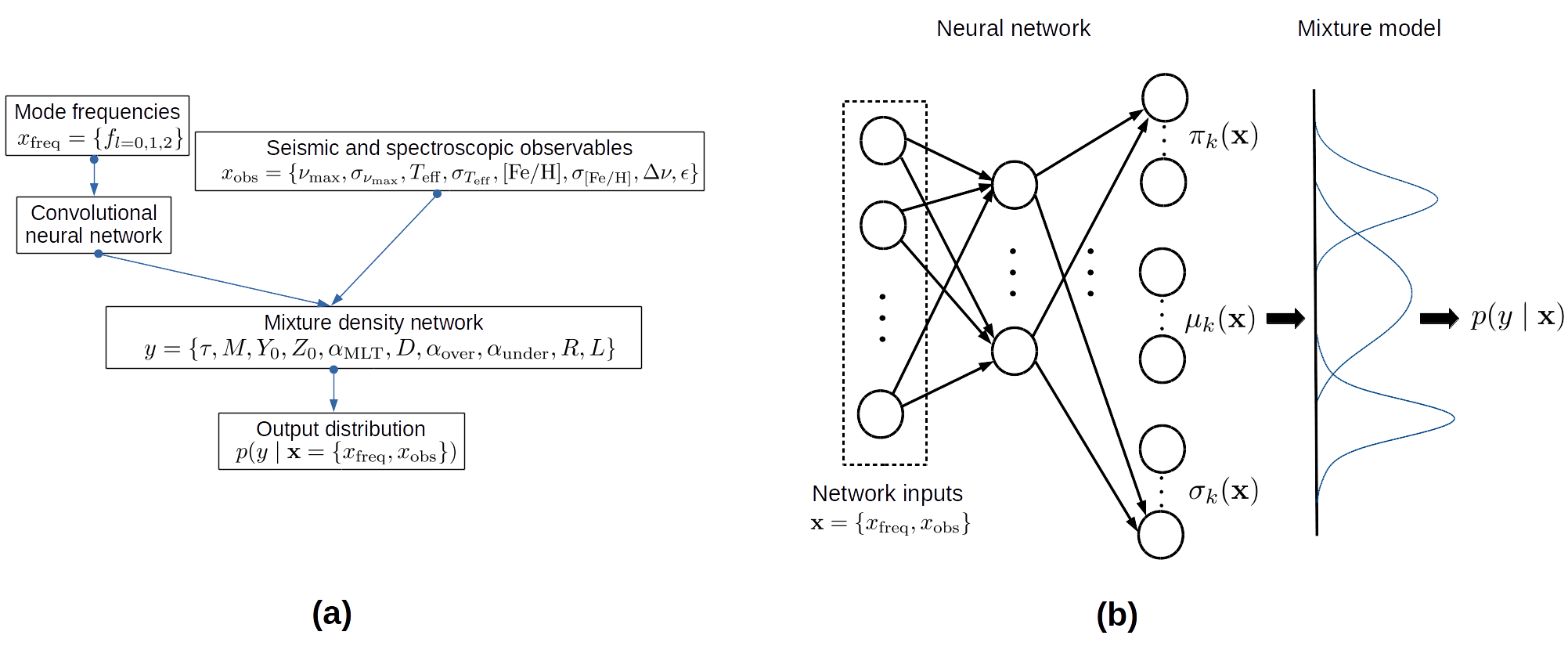}
    \caption{(a) General schematic of the deep neural network in this work. The network takes as input individual mode frequencies, $x_{\mathrm{freq}}$, along with measurements from global seismic parameters and spectroscopic measurements, $x_{\mathrm{obs}}$, to predict the parameters describing a 10-dimensional Gaussian mixture distribution of stellar model parameters, $y$. These parameters are the mean ($\mu$), deviations ($\sigma$), and mixture coefficient ($\pi$) of each Gaussian in the mixture. (b) Schematic of a mixture density network. The network maps input $\mathbf{x}$, which is indicated by the neurons within the box, into conditional density $p(y\mid\mathbf{x})$ by predicting the shape parameters $\pi(\mathbf{x}), \mu(\mathbf{x}), \sigma(\mathbf{x})$ for as many as $k$ Gaussian functions, which are combined to form a mixture model as described in Equations \ref{eq:Gaussian_likelihood} and \ref{eq:MDN_components}.}
    \label{fig:combined_schematic}
\end{figure*}

Owing to high-quality photometric observations from the \textit{Kepler} space mission \citep{Borucki_2010}, precise oscillation frequencies have been measured for subgiant stars (e.g., \citealt{Appourchaux_2012}). Such measurements have enabled the fundamental stellar parameters of subgiants to be determined using stellar modelling techniques
(e.g., \citealt{Metcalfe_2010, Creevey_2012, Dogan_2013, Stokholm_2019}). Although only a small number of oscillating subgiants were observed by \textit{Kepler}, this number is expected to be amplified by
NASA's Transiting Exoplanet Survey Satellite (TESS), where at least a few hundred oscillating subgiants are expected to be observed for a year \citep{Campante_2016, Schofield_2019}. There will therefore be further opportunities for studying subgiant stellar structure and evolution along the subgiant branch.


Stellar models are necessary for inferring stellar ages but the task of finding a model that best fits the observables from a star is computationally demanding. Such a task is a non-linear, high-dimensional optimization problem, where the complex relations governing stellar structure and evolution ($\mathcal{E}$) are sensitive to numerous input physical parameters that are being optimized ($\mathcal{P}$) such as the star's mass, initial composition, and mixing parameters. Traditional optimization methods find a best-matching set of parameters ($\mathcal{P}_*$) that best fits the observed properties of a star ($\mathcal{O}$) by solving the following: 
\begin{equation}
    \mathcal{P}_* = \argmin_\mathcal{P}\left(\frac{\mathcal{E}(\mathcal{P}) - \mathcal{O})^2}{\mathcal{U}^2}\right), 
\end{equation}
\noindent where $\mathcal{U}$ is the uncertainty from $\mathcal{O}$. However, as the dimensionality of $\mathcal{P}$ increases, the volume of the parameter space involved in the search increases exponentially. In an attempt to make stellar model searches tractable, traditional optimization methods typically deploy one or more of the following strategies: lowering the model grid density, grid interpolation (e.g., \citealt{Rendle_2019}), or reducing the number of initial model parameters that are explored in the search. Lowering the grid density significantly reduces the number of models required to be generated, but comes at the cost of parameter coverage that may result in finding sub-optimal solutions. Grid interpolation methods mitigate the need for a very fine grid of models; however they still struggle with high computational complexity once additional dimensions are included in the search. A common alternative is to restrict the search to only a few free parameters and use approximations for other initial model parameters. These include the adoption of a solar-calibrated value for the mixing length parameter ($\alpha_{\mathrm{MLT}}$), or the use of the Galactic enrichment relation to estimate the initial helium abundance ($Y_0$) using the initial metal abundance ($Z_0$). These assumptions may lead to underestimated uncertainties and/or systematic errors when inferring stellar properties from models. An additional prohibiting factor in subgiant model searches is the time-consuming calculation of non-radial modes for evolved stars, which makes it expensive for search methods that require either a large grid of models or the on-the-fly calculation of stellar tracks (e.g., \citealt{Metcalfe_2009, Paxton_2013}).


\citet[hereafter BA16]{Bellinger_2016} showed that these problems can be mitigated for main-sequence stars by using machine learning to infer the parameters of stellar models from a given set of observables. Machine learning techniques, once trained, are able to statistically capture the complex relations connecting observations to stellar models at a fraction of the computational cost required for model grid searches. In other terms, machine learning algorithms can learn to approximate the inverse relation $\mathcal{E}^{-1}$ between model parameters and observed data. As a result, such algorithms output maximum likelihood estimates for $\mathcal{P}_*$ by computing $\mathcal{E}^{-1}(\mathcal{O})$. These algorithms have been shown by BA16 to be effective in the systematic age determination of all main-sequence stars within the high-quality \textit{Kepler} LEGACY sample with an age precision closely comparable to those inferred from traditional grid-based optimization methods \citep{Angelou_2017, Bellinger_2019}.

In this work, we seek to extend machine learning-based stellar model inference towards subgiant stars using deep learning.
A major difference between our work and the BA16 study is the type of asteroseismic stellar age proxy used. The observed oscillation frequency ratios $r_{0,2}$, which are known to be sensitive towards core hydrogen abundance \citep{Roxburgh_2003}, are typically used as a stellar age proxy for main-sequence stars (e.g., \citealt{JCD_1984, White_2011, Bellinger_JCD_2019}). These ratios, however, are no longer effective age proxies for core hydrogen-depleted subgiant stars. Instead, observations of rapidly evolving mixed modes can be used to precisely constrain subgiant stellar ages. The mixed-mode frequency pattern can be analytically described by fitting individual avoided crossings (e.g., \citealt{Deheuvels_2009}), however such an approach can be challenging to compute systematically across a large grid of models that contain both less-evolved and highly-evolved subgiant stars. 
Alternatively, the asymptotic relation of mixed modes \citep{Shibahashi_1979} can be fit to the mixed-mode pattern; however this approach works best for sufficiently evolved subgiants whose coupled g-modes are within the asymptotic regime. Another useful approach to parameterizing mixed modes is with an asteroseismic $p$--$g$ diagram, which shows avoided crossing frequencies versus the $p$-mode large separation as a method to paramaterize subgiant evolution \citep{Bedding_2014}. While useful for a preliminary comparison with theoretical models, extracting precise age estimates with this method would still require detailed modelling of the avoided crossing frequencies. In our work, we introduce a novel machine learning-based method that learns mixed-mode patterns from the \`echelle diagram \citep{Grec_1983} and therefore does not require such patterns to be explicitly parameterized. As a result, our method can estimate the ages of oscillating stars from early post-core hydrogen exhaustion up to the base of the red-giant branch.

While machine learning has previously been applied for asteroseismic modelling (e.g., \citealt{Verma_2016, Bellinger_2016, Hendriks_2019}), another novelty in our approach is the estimation of parameters in the form of distributions, rather than point estimates. Our method estimates a distribution across an 8D parameter space with relatively small computational cost. Besides five basic input model parameters, namely age ($\tau$), mass ($M$), initial fractional helium abundance ($Y_0$), initial fractional metal abundance ($Z_0$) and mixing length parameter ($\alpha_{\mathrm{MLT}}$), we include additional processes in the form of convective core overshooting, envelope undershooting, and heavy element diffusion. These processes have their respective free parameters in the form of the overshooting parameter ($\alpha_{\mathrm{over}}$), undershooting parameter ($\alpha_{\mathrm{under}}$), and diffusion multiplication factor ($D$), all of which have complex influences on the evolution of a subgiant star. For instance, the $\alpha_{\mathrm{over}}$ alters the size of a star's convective core on the main sequence. Not only does this affect the amount of fuel the star has to prolong its main sequence lifetime, but it also changes the core's central density at a certain age as a subgiant \citep{Deheuvels_2011}. A similar effect is achieved with the coefficient $D$ that controls the effect of microscopic diffusion in low-mass stars: the process sinks heavy elements while dispersing hydrogen towards the surface, which reduces a star's age at a given mean density (e.g., \citealt{Miglio_2005, Gai_2009, Valle_2015}). Meanwhile, the undershooting parameter $\alpha_{\mathrm{under}}$ controls the inwards extent of the outer convective boundary of the stellar envelope and is often constrained to be equivalent with $\alpha_{\mathrm{over}}$. For exploratory purposes, we set $\alpha_{\mathrm{under}}$ to be a free parameter. 

Despite much evidence in literature indicating the importance of these additional processes in stellar models (e.g., \citealt{Guzik_1993, Gruyters_2013, SilvaAguirre_2013}), there remains significant uncertainty in both theory and observations regarding the nature and efficiency of such processes. It is therefore common for modelling tasks to either disregard the parameters governing these additional processes as free parameters in a grid of models or to generate multiple grids to test different fixed levels of efficiency for these additional processes \citep{Silva_Aguirre_2015}. 
By including the parameters governing these additional processes within the grid of models in our study, we explore a wider range of solutions for subgiant fundamental parameters with minimal assumptions about the input physics\footnotemark of the grid. The use of machine learning to estimate additional input physics parameters from the grid additionally opens up possibilities of empirically estimating relations between model parameters such as the $M-\alpha_{\mathrm{over}}$ relation \citep{Angelou_2020} or the $\alpha_{\mathrm{MLT}}$-[Fe/H] relation \citep{Viani_2018}.

Our work in this study is expected to form an efficient method for subgiant star fundamental parameter estimation that will enable the characterization of subgiant ensembles and support the grid-based modelling of individual subgiant stars by providing informative estimates. First, we detail the construction of our deep learning algorithm and a novel sampling-based training procedure to increase the network's robustness towards measurement uncertainties and known systematics in stellar models. We then report the performance of our method on a hold-out set of subgiant stellar models and estimate the properties of real subgiant stars, which includes those modelled individually as well as those modelled as part of an ensemble.\footnotetext{Although processes like convection, convective overshoot, or microscopic diffusion are typically approximated only by empirical treatments, such treatments are commonly referred to as `input physics' within grid-based modelling studies. Our use of the term `additional input physics parameters' in this work thus refers to parameters $\alpha_{\mathrm{over}}$, $\alpha_{\mathrm{under}}$, and $D$ that govern the treatments of convective overshooting/undershooting and microscopic diffusion, respectively.}

\begin{table}
    \centering
    \begin{tabular}{llll}
    \hline
        Symbol & Name & Min & Max \\\hline
        $M/$M$_\odot$ & Mass & $0.7$ & $1.8$ \\
        $Y_0$ & Fractional helium abundance & $0.22$ & $0.34$ \\
        $\alpha_{\text{MLT}}$ & Mixing length parameter & 1 & 3 \\
        $Z_0$ & Fractional metal abundance & $0.0001$ & $0.04$ \\
        $\alpha_{\text{over}}$ & Overshooting parameter & $0.0001$ & 1 \\
        $\alpha_{\text{under}}$ & Undershooting parameter & $0.0001$ & 1 \\
        $D$ & Diffusion multiplication factor & $0.0001$ & 3\\\hline
    \end{tabular}
    \caption{Ranges of initial parameters in the computed grid of stellar evolution models. The latter four parameters are varied logarithmically, and the latter three values are set to 0 if their value would otherwise be less than $0.001$. \label{tab:param-ranges}} 
\end{table}

\section{Method} \label{method}
We develop a deep neural network that predicts $\tau$, $M$, $Y_0$, $Z_0$, $\alpha_{\mathrm{MLT}}$, $D$, $\alpha_{\mathrm{over}}$, and $\alpha_{\mathrm{under}}$ of oscillating subgiant stars. We additionally estimate stellar radius ($R$) and luminosity ($L$), thus increasing the dimensionality of the network's output to ten. The network takes as input individual mode frequencies, the global seismic parameter $\nu_{\mathrm{max}}$, and spectroscopic observables ($T_{\mathrm{eff}}$, [Fe/H]). We train the network with supervised learning on a grid of models that we describe in Section \ref{model_grid}. In Section \ref{DNN}, we detail the deep neural network's structure and training procedure.

\subsection{Models for Training}\label{model_grid}
We use \emph{Models for Experiments in Stellar Astrophysics} \citep[\textsc{Mesa}~r12778,][]{Paxton_2011, Paxton_2013, Paxton_2015, Paxton_2018, Paxton_2019} to compute a dense grid of stellar models. 
The calculations begin at the pre-main-sequence evolutionary phase and span until the base of the red-giant branch. 
The set of input physics of the evolution is the same as described in BA16 and \citet{Bellinger_2019}. 
The input parameters of each track ($M, Y_0, Z_0, \alpha_{\text{MLT}}, \alpha_{\text{under}}, \alpha_{\text{over}},$ and $D$) are varied quasi-randomly (see Appendix B of BA16) in the ranges listed in Table~\ref{tab:param-ranges}. 

As in \citet{Bellinger_2019}, we define three evolutionary stages of interest: the main-sequence (MS), the MS turn-off (TO), and subgiant branch (SG). 
We define the beginning of the MS as when at least $99.99\%$ of the stellar luminosity is generated by hydrogen fusion. 
We define the beginning of the TO (and end of MS) as the point when the central hydrogen abundance ($X_c$) drops below $10^{-1}$. 
We define the beginning of the SG branch (and end of TO) as the point when $X_c$ drops below $10^{-6}$.
Finally, we define the end of the SG branch as when ${\text{d} \log L / \text{d} \log T_{\text{eff}} > -3}$ or the asymptotic period spacing drops below $150$ seconds, whichever happens first. The latter condition is in accordance with the period spacing at the end of the subgiant phase as measured by \citet{Mosser_2014}. Alternatively, any phase can end if a maximum age of $15$~Gyr is reached, after which no subsequent phases are computed. 

From each of these phases, we retain $32$ models which we select to be nearly equally spaced (see Appendix A of BA16) either in $X_c$ (in the case of MS models) or in age (for TO and SG models). 
We use \textsc{Gyre} \citep{Townsend_2013, Townsend_2018} to compute the radial (spherical degree ${\ell = 0}$) and non-radial (${1 \leq \ell \leq 3}$) linear adiabatic mode frequencies and inertias of these models. 
In total, the grid contains 660,736 stellar models. For our training set in this study, we select models with ($X_c < 10^{-5}$), resulting in 271,631 models near the end of the TO phase up to the end of the SG branch.

\subsection{Neural Network}\label{DNN}
The deep neural network, as visualized in Figure \ref{fig:combined_schematic}a, comprises two components: a convolutional neural network and a mixture density network. The detailed structure of the full network is described in Appendix \ref{Appendix:Network_Architecture}, and the code for performing estimates and training a network is made available at \url{https://github.com/mtyhon/deep-sub}.  In the following, we describe the role of each network component.






\subsubsection{Convolutional Neural Network: Analyzing Oscillation Modes}
\begin{figure}
    \centering
	\includegraphics[width=1.\columnwidth]{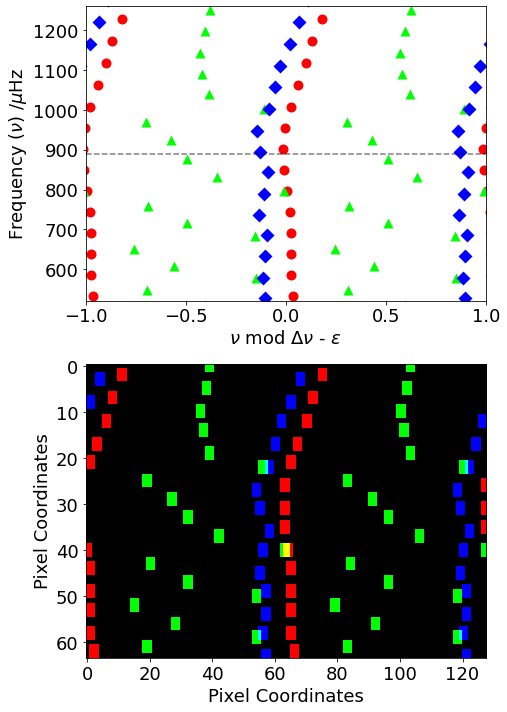}
	\vspace{-15pt}
    \caption{(Top) A repeated \'echelle diagram of a subgiant model's oscillation spectrum showing $l=1$ avoided crossings. $l=0$ modes are represented as red circles, $l=1$ modes as green triangles, and $l=2$ modes as blue diamonds. The diagram's vertical axis has a range of $\pm 7 \Delta\nu$ around $\nu_{\mathrm{max}}$ (dashed line). Additionally, the oscillation modes are positioned such that the $l=0$ ridge aligns with $\epsilon$ calculated from the 6 closest $l=0$ modes to $\nu_{\mathrm{max}}$. (Bottom) The same \'echelle diagram binned into a 128x64 image as input for the convolutional neural network. Each $l=0$ (red), $l=1$ (blue), and $l=2$ (green) mode occupies a 5x5 square within the image. The use of three separate colour channels allows overlapping modes in the image to still be visible to both viewer and network.}
    \label{fig:echelle}
\end{figure}
The role of the convolutional neural network in our method is to detect the mixed-mode patterns from oscillation modes by automatically learning pattern-matching filters from training data. Because we want to emphasize both the near-uniform regularity of p modes as well as the mixed-mode pattern, the oscillation modes are represented in a repeated \'echelle diagram that is provided as input to the network in the form of a 2D image as shown in Figure \ref{fig:echelle}. 
The advantages of such a representation are as follows:

\begin{itemize}
    \item An \'echelle diagram distinctly shows the mixed-mode pattern without requiring the detailed parameterization of each mode frequency or avoided crossing.
    \item The network can easily adapt to missing oscillation modes that can occur for low S/N observations. Because most data-driven methods require their inputs to have a fixed size, we can only use a fixed number of modes per star/model if we use numerical frequency values as the input. This is circumvented by using an \'echelle diagram because the size of the 2D image of the diagram remains constant regardless of the number of modes present.
    \item Due to the binning of mode frequencies as a 128x64 input image, the input to the network is unchanged in the presence of relatively small frequency shifts. In particular, the position of a mode in the image will only shift vertically if the mode is perturbed with a frequency magnitude of at least $7\Delta\nu$/64 $\mu$Hz. Shifting a mode horizontally in the diagram would require a frequency perturbation of at least $2\Delta\nu$/128 $\mu$Hz. Assuming a subgiant $\Delta\nu$ of $\sim50\;\mu$Hz, a horizontal mode shift would require a frequency perturbation of at least $\sim0.75\;\mu$Hz, which is typically at the 3$\sigma$ level for frequency measurements in \textit{Kepler} data. The binning of mode frequencies thus encompasses the uncertainty in frequency measurements and prevents the network from overfitting.
\end{itemize}

To create the \'echelle diagram of a given model, we first estimate the frequency of maximum oscillation power, $\nu_{\mathrm{max}}$, using the following scaling relation \citep{Brown_1991, Kjeldsen_1995}: 

\begin{equation}
    \nu_{\mathrm{max}} = \frac{M/M_{\odot}}{(R/R_{\odot})^2\sqrt{T_{\mathrm{eff}}/T_{\mathrm{eff}, \odot} }} \nu_{\mathrm{max}, \odot},
\end{equation}
with $\nu_{\mathrm{max}, \odot} = 3090\;\mu$Hz \citep{Huber_2011} and $T_{\mathrm{eff, \odot}}=5772$~K \citep{Pra_2016}. Using the 6 nearest $l=0$ modes to $\nu_{\mathrm{max}}$, we calculate $\Delta\nu$ and the offset $\epsilon$ using a weighted linear fit to the following equation:

\begin{equation}\label{eq:asymptotic}
    \nu = \Delta\nu(n + \epsilon),
\end{equation}
where $n$ is the mode order and $\nu$ is the mode frequency. Note that we define the offset to be $\epsilon =(\nu/\Delta\nu)$ modulo 1. Therefore, the exact value of $n$ is not required as long as the radial modes are correctly ordered by a spacing of $\Delta\nu$ \footnote{The true value of the offset varies smoothly between a range of $0.6\apprle\epsilon\apprle1.6$ as a subgiant evolves (e.g., \citealt{White_2011}). For the ease of implementation in our algorithm, our definition of $\epsilon =(\nu/\Delta\nu)$ modulo 1 bounds the offset between 0 and 1.}.

 The fit is weighted by a Gaussian centered at $\nu_{\mathrm{max}}$ with a standard deviation of $0.1$$\nu_{\mathrm{max}}$. 

Next, we construct a repeated \'echelle diagram with a range of $\pm\Delta\nu$ on the horizontal axis and $\pm7\Delta\nu$ around $\nu_{\mathrm{max}}$ on the vertical axis. We additionally shift the abscissa of the \'echelle diagram by $\epsilon$ such that an $l=0$ ridge is always positioned at the center of the diagram. Finally, we bin the diagram into a 2D array of size 128x64 as input to the network.


\subsubsection{Mixture Density Network}
After the pattern analysis of mode frequencies with the convolutional neural network, a mixture density network \citep[MDN]{Bishop_1994} combines the mode frequency information with other spectroscopic and global seismic parameters.
Given the network input $\mathbf{x}$, a MDN models the conditional density $p({y\mid\mathbf{x}})$ of the output parameter vector ${y = \{\tau, M, Y_0, Z_0, \alpha_{\mathrm{MLT}}, D, \alpha_{\mathrm{over}}, \alpha_{\mathrm{under}}, R, L\}}$ as a Gaussian mixture model that is given by the following:
\begin{equation}\label{eq:Gaussian_likelihood}
    p(y\mid \mathbf{x}) = \sum_{k=1}^K\pi_k(\mathbf{x}) \frac{1}{(2\pi)^{N/2}\sigma_k(\mathbf{x})}\exp{\left(- \frac{(y-\mu_k(\mathbf{x}))^2}{2\sigma_k(\mathbf{x})^2}\right)},
\end{equation}
where ${N=10}$ is the number of output parameters and $K$ is the number of Gaussian distributions. Each distribution is parameterized with a mean value of $\mu_k(\mathbf{x})$, a shape factor of $\sigma_k(\mathbf{x})$, and a mixing coefficient of $\pi_k(\mathbf{x})$. For our study, we specify the network output to be described by as many as $K=16$ distributions\footnote{$K=16$ was determined to yield the lowest negative log-likelihood (Equation \ref{eq:nll}) without overfitting the data. More details are shown in Appendix \ref{Appendix:num_gaussians}}. The MDN output for each parameter is a vector $w$ of length $3K$, comprising the following:
\begingroup
 \large
\begin{equation}
  \begin{array}{l}
    \mu_k(\mathbf{x}) = w_k^\mu,\\
    \sigma_k(\mathbf{x}) = w_k^\sigma,\\

    \pi_k(\mathbf{x}) = \frac{\exp({w_k^\pi})}{\sum_{k=1}^K\exp({w_k^\pi})},
  \end{array}\label{eq:MDN_components}
\end{equation}
\endgroup
with $k\in [1,...,K]$ and $\mu_k$, $\sigma_k$, and $\pi_k$ representing the respective mean, standard deviation, and mixing coefficient of the $k-$th mixture component with $\sum_{k=1}^K \pi_k(\mathbf{x}) = 1$. A schematic of the MDN is shown in Figure \ref{fig:combined_schematic}b. Because there are 10 parameters that are estimated by the MDN, both $\mu_k$ and $\sigma_k$ in this work are 10-dimensional. Optimizing the MDN during training involves minimizing the negative log-likelihood $E$, given by the following:

\begin{equation}\label{eq:nll}
    E = \sum_{m=1}^{m_\mathrm{tot}}-\ln{p(y_m\mid \mathbf{x}_m)},
\end{equation}
where $m_\mathrm{tot}$ is the total number of models in the training set.

Fundamentally, we expect each output parameter in $y$ to span a distribution within a grid of stellar models when given a set of subgiant star observables $\mathbf{x}$, which is why conditional density estimation with an MDN is useful. The MDN's output is effectively a region of parameter space that is expected to contain the global optimum, with uncertainties that can be estimated directly from the properties of the output parameter distribution. This is a highly efficient way of obtaining good initial guesses spanning a narrow region of parameter space for traditional grid optimization approaches. Additionally, output estimates in the form of distributions express more explicitly the presence of non-unique solutions within a grid of models, which are often the largest sources of uncertainty in subgiant star model fitting (e.g., \citealt{Dogan_2013}). For instance, minor adjustments to the input physics of subgiant stellar models can cause them to share the same luminosity even though they have different masses, as discussed by \citet[their Section 5.2]{Metcalfe_2010}.


    


\subsection{Training the Network} \label{Network_Training}
The network is trained over 500 iterations, with early stopping if the network's performance on a hold-out validation set does not improve after more than 20 consecutive iterations. Network training only incurs a one-time cost of 2-3 hours using an NVIDIA Titan Xp GPU. Once trained, estimating the properties for a subgiant is extremely efficient, typically requiring less than one second per star.

During training, we perform bootstrapping of the input data, meaning that the values we pass to the network for each training iteration are randomly perturbed by noise or by artificially-included systematic offsets. The goal with bootstrapping is to
train the network to recover the correct model values even when they have been perturbed by noise or systematic offsets. At the same time, it prevents the network from overfitting on the grid of models. The following sections describe each step in the bootstrapping procedure, with an outline in the form of pseudo-code presented in Appendix \ref{Appendix_Bootstrapping}.

\subsubsection{Surface Correction}
The improper modelling of the near-surface layers in 1D stellar models results in a systematic offset of model frequencies from the mode frequencies of real solar-like oscillators. This frequency offset is known as the \textit{surface effect}, which varies proportionally with the inverse of mode inertia \citep{Gough_1990}. A correction term to the surface effect, $\delta\nu_{\mathrm{surf}}$, was proposed by \citet{Ball_2014} and is given by the following equation:
\begin{equation}
    \delta\nu_{\mathrm{surf}} = [c\cdot(\nu/\nu_{\mathrm{ac}})^3 + a\cdot(\nu/\nu_{\mathrm{ac}})^{-1}]/\mathcal{I}, \label{cubic_equation}
\end{equation}
where $\nu$ is the mode frequency, $\nu_{\mathrm{ac}}$ is the acoustic cut-off frequency, $\mathcal{I}$ is the normalized mode inertia, and both $c$ and $a$ are coefficients that are determined by matching the model frequencies to the observed frequencies. 

When training the network, we randomly apply different levels of surface term corrections to all model frequencies. For implementation simplicity, we use only the cubic term in Equation \ref{cubic_equation} and determine for each model the range of parameter $c$ required to obtain a $\delta\nu_{\mathrm{surf}}$ between 0.22-0.38\% of $\nu_{\mathrm{max}}$ for the $l=0$ mode closest to $\nu_{\mathrm{max}}$. This $\delta\nu_{\mathrm{surf}}$ range is empirically estimated based on frequency offsets reported by \citet{Ball_2017} for subgiant stars. Each stellar model in the training set thus has its own uniform range of values that $c$ can take. In every training iteration, we randomly sample $c$ for each model, calculate their corresponding $\delta\nu_{\mathrm{surf}}$, and offset each model's oscillation frequencies to simulate the frequencies from a real star. Because $c$ for each model is randomly sampled in every training iteration, different levels of surface term offsets are consistently simulated during training. By covering the range of variations expected for $\delta\nu_{\mathrm{surf}}$, we aim to increase the network's robustness towards the surface effect.


\subsubsection{Frequency Perturbation}\label{section:freq_perturbation}
Besides an artificial correction to the surface term, the input model frequencies are perturbed with random noise during training. The $l=0$ modes of each stellar model are perturbed by Gaussian noise with a standard deviation of $\sigma_{l=0}$. The value of $\sigma_{l=0}$ is uniformly sampled from a range of 0.1-1$\mu$Hz in each training iteration. $l=1$ and $l=2$ modes for each stellar model are also perturbed with noise, but with $\sigma_{l=1} = (0.5-1)\;\sigma_{l=0}$ and $\sigma_{l=2} = (1-2)\;\sigma_{l=0}$, which are estimated from the relative uncertainties of mode frequencies for main sequence stars in the \textit{Kepler} LEGACY sample \citep{Lund_2017}. Compared to the $l=1$ modes of main sequence stars, the mixed $l=1$ modes of subgiants have larger inertiae and subsequently smaller observed linewidths due to the increased mode coupling between core and envelope (e.g., \citealt{Grosjean_2014}).
While this indicates that our implementation may overestimate the uncertainties of mixed $l=1$ modes, we choose to be conservative with our uncertainties.

\subsubsection{Simulating Missing Modes}
For lower S/N observations of subgiant stars, it is common to have individual modes missing within oscillation spectra. To train our network to be robust towards this phenomenon, we randomly remove modes from the \'echelle diagram in each training iteration. The number of modes retained in the \'echelle diagram is dependent on $l$: we retain $l=0$ modes within a $4-7\;\Delta\nu$ range from $\nu_{\mathrm{max}}$, while $l=1$ modes are retained in a similar but independent manner from the $l=0$ modes. Meanwhile, the $\Delta\nu$ range for retained $l=2$ modes are constrained to be smaller or equal to the model's $l=0$ range.
In addition to varying the range of oscillation modes, we apply a 5\% chance for each mode to be randomly removed from the set of model frequencies.


\begin{table} 
	\centering
	\caption{Summary of perturbation magnitudes of network inputs.}
	\label{PerturbationMagnitudeTable}
		\begin{tabular}{|c|c|}
			\hline
			Input & Perturbation magnitude \\				
			\hline
			$\nu_{\mathrm{max}}$ & 0.5$-$10\% \\
			$T_{\mathrm{eff}}$ & $50-150$~K\\
			$[$Fe/H$]$ & $0.05-0.15$~dex \\
			$\sigma_{l=0}$ & $0.1-1\;\mu\mathrm{Hz}$\\
			$\sigma_{l=1}$ & $(0.5-1)\;\sigma_{l=0}$\\
			$\sigma_{l=2}$ & $(1-2)\;\sigma_{l=0}$\\
			\hline
		\end{tabular}
\end{table}

\subsubsection{Noise in Spectroscopic and Global Seismic Parameters}
Similar to the frequency perturbations in Section \ref{section:freq_perturbation}, we perturb the $\nu_{\mathrm{max}}$, $T_{\mathrm{eff}}$, and [Fe/H] values of each model with random Gaussian noise so that the network learns to recover model values in the presence of noisy spectroscopic and global seismic parameters. In each training iteration, the magnitudes of $\sigma_{\nu_{\mathrm{max}}}$, $\sigma_{T_{\mathrm{eff}}}$, and  $\sigma_{\mathrm{[Fe/H]}}$ describing the Gaussian noise are sampled uniformly from a range of values as in Table \ref{PerturbationMagnitudeTable}.

\section{Results}\label{results}
\subsection{Validation Set} \label{test_set_performance}

\begin{table} 
	\centering
	\caption{Validation metrics on a hold-out set of stellar models. The metrics reported are the mean absolute percentage error (MAPE), mean absolute error (MAE), and the explained variance score $V$ (Equation \ref{eq:explained_variance}). These metrics assume the use of only a point estimate (the distribution mean) to quantify performance, and thus low performance values for a parameter implies that its distribution is non-localized in parameter space.  }\label{table:metrics}
	\label{PerformanceTable}
		\begin{tabular}{|c|c|c|c|}
			\toprule
			Output Parameter & MAPE & MAE & $V$ \\				
			\midrule
			$\tau$ & 8.12\% & 0.34~Gyr & 0.97\\
			$M$ & 3.40\% & 0.04~$M_{\odot}$ & 0.96\\
			$R$ & 1.10\% & 0.02~$R_{\odot}$ & 0.99\\
			$L$ & 4.73\% & 0.64~$L_{\odot}$ & 0.99\\
			$Y_0$ & 7.50\% & 0.02& 0.41 \\
			$Z_0$ & 16.8\% & 0.01& 0.96\\
			$\alpha_{\mathrm{MLT}}$ & 15.8\% & 0.27 & 0.53\\
			$\alpha_{\mathrm{over}}$ & 120\% & 0.08 & 0.10 \\
			$\alpha_{\mathrm{under}}$ & 145\% & 0.14 & -0.09\\
			$D$ & 133\% & 0.30 & 0.18\\
			\bottomrule
		\end{tabular}
\end{table}

\begin{figure}
    \centering
	\includegraphics[width=0.98\columnwidth]{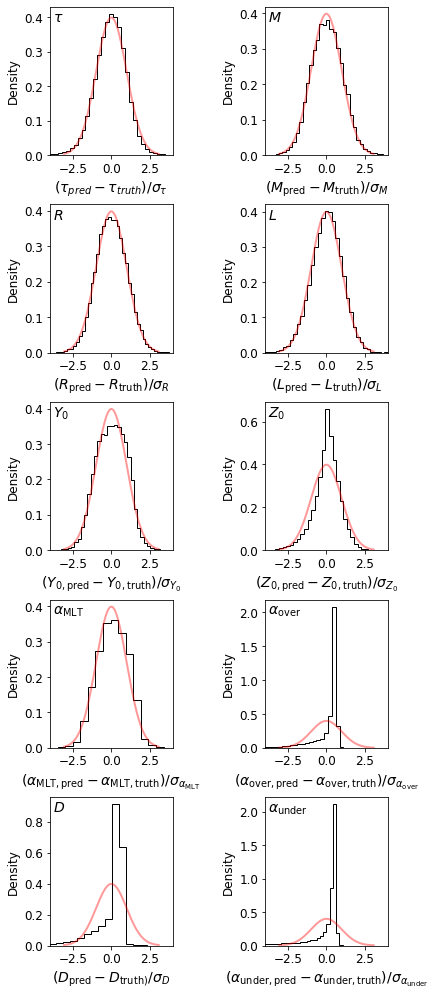}
    \caption{The z-score distribution for the estimated distribution mean for each output parameter in the test set. The value of $\sigma$ for each parameter is calculated as the square root of its estimated distribution's total variance. Because each distribution comprises a superposition of $k$ Gaussian distributions with mean $\mu_k$ and deviation $\sigma_k$, the total variance is calculated by adding the expectation of $\sigma_k^2$ to Var($\mu_k$), i.e. the Law of Total Variance.
   When the z-score is normally distributed (red), the estimated distribution mean on average has no systematic offsets from the true value and the estimated $\sigma$ neither overestimates nor underestimates the true uncertainties. The unique distributions for $\alpha_{\mathrm{over}}, \alpha_{\mathrm{under}}$, and $D$ are caused by such parameters being not highly constrained with large uncertainties (see text).}
    \label{fig:z_statistic}
\end{figure}

\begin{table*} 
	\centering
	\begin{threeparttable}
	\caption{Network inputs for \textit{Kepler} subgiants that have been analyzed individually using asteroseismic grid-based modelling. Unless specified otherwise, the spectroscopic parameters and $\nu_{\mathrm{max}}$ for a star are from the same source study as the mode frequencies. } \label{Table:NetworkInputs}
	\label{table:Network}
		\begin{tabular}{|l|l|l|l|l|}
			\toprule
			Star & Mode frequencies & $T_{\mathrm{eff}}$~(K) & [Fe/H]~(dex) & $\nu_{\mathrm{max}}$~($\mu$Hz)\\
			\midrule
			Gemma & From \citet{Appourchaux_2012} & $5682\pm84^{a}$ & $0.05\pm0.09^{a}$ & $890\pm12^{b}$\\
			Scully & Maximal set from \citet{Campante_2011} & $5790\pm74^{c}$ & $-0.04\pm0.10^{c}$ & $990\pm60^{c}$\\
			Boogie & Maximal set from \citet{Mathur_2011} & $5700\pm100^{c}$ & $0.13\pm0.10^{c}$ & $847\pm16^{b}$\\
			HR 7322 & From \citet{Stokholm_2019} & $6313\pm50$ & $-0.23\pm0.06$ & $960\pm15$\\
			\bottomrule
		\end{tabular}
		\begin{tablenotes}
      \small
      \item $^a$ \citet{Bruntt_2012}
      \item $^b$ \citet{Serenelli_2017}
      \item $^c$ \citet{Creevey_2012}
    \end{tablenotes}
  \end{threeparttable}
\end{table*}

\begin{table*}
\setlength\tabcolsep{2.25pt}
	\centering 
    \caption{Estimates for \textit{Kepler} subgiants that have been analyzed individually using asteroseismic grid-based modelling. For each parameter, the quoted uncertainties from this work represent the 16th and 84th percentile values. The estimated probability densities for each star are shown in Appendix \ref{Appendix:10d}.}\label{table:SummaryGridBasedModelling}
    \begin{tabular}{|l|l|c|l|c|l|c|l|c|}
    
    \toprule
		    & \multicolumn{2}{c|}{Gemma}& \multicolumn{2}{|c|}{Scully}& \multicolumn{2}{|c|}{Boogie}& \multicolumn{2}{|c|}{HR 7322}\\
			\cmidrule(lr){2-3}
			\cmidrule(lr){4-5}
			\cmidrule(lr){6-7}
			\cmidrule(lr){8-9}
			& This work & \citet{Metcalfe_2014}& This work & \citet{Dogan_2013} & This work & \citet{Dogan_2013} & This work & \citet{Stokholm_2019}\\
			\midrule
			
			$\tau$~(Gyr) & $4.92^{+0.64}_{-0.40}$ & $5.00\pm0.53$ & $6.17^{+1.02}_{-0.88}$ & $7.12\pm0.47$& $4.45^{+0.57}_{-0.38}$ & $4.57\pm0.23$& $3.32^{+0.32}_{-0.21}$ & $4.27^{+0.05}_{-0.04}$\vspace{0.075cm}\\
			$M$~($M_{\odot}$) &  $1.23\pm0.04$ & $1.27\pm0.06$ &  $1.13\pm0.05$ & $1.00\pm0.04$&  $1.32^{+0.05}_{-0.06}$ & $1.27\pm0.04$&  $1.30\pm0.04$ & $1.20\pm0.01$\vspace{0.075cm}\\
			$R$~($R_{\odot}$)& $2.086^{+0.027}_{-0.024}$ & $2.106 \pm 0.025$ & $1.857^{+0.027}_{-0.025}$ & $1.776 \pm 0.021$& $2.201^{+0.029}_{-0.033}$ & $2.184\pm0.024$& $2.008^{+0.021}_{-0.022}$ & $1.954 \pm 0.006$\vspace{0.075cm}\\
			$L$~($L_{\odot}$) & $4.05^{+0.27}_{-0.25}$ & $4.17\pm0.27$ & $3.60^{+0.26}_{-0.24}$ & $3.18\pm0.13$& $4.58^{+0.33}_{-0.31}$ & $4.54\pm0.30$& $5.72\pm0.25$ & $5.37\pm0.06$\vspace{0.075cm}\\
			$Y_0$ & $0.261^{+0.030}_{-0.026}$ & $0.254 \pm 0.016$ & $0.283^{+0.030}_{-0.028}$ & $0.294 \pm 0.014$& $0.256^{+0.032}_{-0.024}$ & $0.276 \pm 0.022$& $0.248^{+0.028}_{-0.019}$ & $0.261 \pm 0.001$\vspace{0.075cm}\\
			$Z_0$ & $0.019^{+0.003}_{-0.002}$ & $0.020 \pm 0.003$ & $0.019\pm0.003$ & $0.011 \pm 0.002$& $0.023^{+0.004}_{-0.002}$ & $0.023 \pm 0.003$& $0.011^{+0.003}_{-0.001}$ & $0.010 \pm 0.001$\vspace{0.075cm}\\
			$\alpha_{\mathrm{MLT}}$ & $1.80^{+0.10}_{-0.09}$ & $2.10 \pm 0.37$& $1.96^{+0.14}_{-0.11}$ & $1.96 \pm 0.09$& $1.90^{+0.13}_{-0.11}$ & $1.91 \pm 0.09$& $1.83\pm0.09$ & 1.60 \vspace{0.075cm}\\
			$\alpha_{\mathrm{over}}$ & $0.007^{+0.115}_{-0.006}$ & - &  $0.021^{+0.322}_{-0.020}$ & - & $0.009^{+0.158}_{-0.008}$ & - & $0.006^{+0.077}_{-0.005}$ & - \vspace{0.075cm}\\
			$\alpha_{\mathrm{under}}$ & $0.005^{+0.246}_{-0.004}$ & -& $0.004^{+0.088}_{-0.003}$ & -& $0.005^{+0.302}_{-0.004}$ & -& $0.010^{+0.352}_{-0.009}$ & - \vspace{0.075cm}\\
			$D$ & $0.024^{+0.765}_{-0.023}$ & - & $0.105^{+1.147}_{-0.104}$ & -& $0.022^{+0.925}_{-0.021}$ & -& $0.017^{+0.661}_{-0.016}$ & -\\
			\bottomrule
		\end{tabular}
\end{table*}

To quantify how well the network can recover parameters from our grid of models, we measure its performance on a test set comprising 995 tracks from the grid that were not used for training. For each output estimate comprising a mixture of $k$ Gaussian distributions, we take the predicted value $\mathbf{\hat y}$ to be the sum of each distribution's mean, weighted by $\pi_k$. We report the following metrics between $\mathbf{\hat y}$ and the true model values $\mathbf{y}$: the mean absolute error (MAE), the mean absolute percentage error (MAPE), and the explained variance score. The explained variance score is defined by the following:
\begin{equation}\label{eq:explained_variance}
    V = 1-\frac{\mathrm{Var}(\mathbf{y}-\mathbf{\hat y})}{\mathrm{Var}(\mathbf{y})},
\end{equation}
with Var indicating the variance. 
This metric measures how well the network captures the variance of an output parameter in the test set, and ranges between negative infinity in the worst case scenario; and one for a perfect predictor. Meanwhile, the MAPE and MAE measure how well the estimated distribution's mean (a point estimate) can approximate the true model value.  These metrics are tabulated in Table \ref{PerformanceTable}, and are further discussed in Section \ref{section:validation_interpretation}. Besides performance metrics, we additionally evaluate the quality of our predicted uncertainties by visualizing each output parameter's z-score, defined as $(\mathbf{y} - \mathbf{\hat y})/\sigma_{\mathbf{\hat y}}$. 
Each parameter's z-score over the validation set is shown in Figure \ref{fig:z_statistic}, where in each panel a comparison is made to a normal distribution (plotted in red). 
The skewness of the z-score relative to a normal distribution indicates an average systematic offset between predicted and true values in the test set. Furthermore, the increased or decreased sharpness of the z-score relative to a normal distribution indicates underestimated or overestimated uncertainties, respectively.

\subsection{Interpretation of Validation Results}\label{section:validation_interpretation}
The analysis in Section \ref{test_set_performance} indicates how well a point estimate in the form of the distribution mean of each output parameter can match the true model value. If a parameter distribution is broad or multi-modal, the distribution mean becomes imprecise, resulting in larger MAPE and MAE, and a smaller $V$. The validation metrics as described in Table \ref{table:metrics} therefore shows how well the input (comprising asteroseismic and spectroscopic measurements) can constrain each output parameter. For instance, having mass ($M$) and radius ($R$) as the most precisely estimated parameters indicates that subgiant masses and radii are highly constrained to a narrow parameter range that can be approximated well using the mean of their corresponding estimated distributions. Such a result is expected given that $\Delta\nu$, $\nu_{\mathrm{max}}$, and $T_{\mathrm{eff}}$ \textemdash{} all of which are parameters that can be used to infer mass and radii using the asteroseismic scaling relations \citep{Brown_1991,Kjeldsen_1995} \textemdash{} are provided as inputs to the network. Stellar ages, $\tau$ are well-constrained with an average error of 8\%. The similarity of the z-score distribution to a normal distribution for parameters $M$, $R$, and $\tau$ demonstrates that on average, the reported uncertainties for these parameters correctly reflect the deviation of the estimated mean from the true value.

Parameters $Y_0$ and $\alpha_{\mathrm{MLT}}$ are only moderately constrained and thus show some degeneracy in their values. This means that over a moderate range, such parameters can have multiple combinations that provide good matches to a subgiant's observables (e.g., \citealt{Deheuvels_2011}). With a high $V$ of 0.96, $Z_0$ is considered to be well-constrained. Its relatively high MAPE is a consequence of its logarithmic variation throughout the model grid. The z-score distribution for $Z_0$, which is sharper compared to a normal distribution, indicates that $\sigma_{Z_0}$ values are overestimated on average.

Additional input physics parameters $\alpha_{\mathrm{over}}$, $\alpha_{\mathrm{under}}$, and $D$ have high MAPE and low $V$ values in Table \ref{table:metrics} and therefore are not precisely estimated by the estimated distribution mean. The z-score distribution for these parameters show very sharp distributions, indicating large uncertainties regardless of how close the estimated distribution mean is to the model value. These results imply 
that across our high-dimensional grid in this work, each additional input physics parameter can have a broad range of likely values for a given set of input observables.


\subsection{Fundamental Parameter Estimation: Comparison with Classical Grid-based Modelling}\label{Section:Boutique}
To test our method, we apply it to four \textit{Kepler} subgiant stars that have been individually modelled using classical asteroseismic grid-based search techniques, namely KIC 11026764, KIC 10920273 , KIC 11395018, and KIC 10005473. The first three stars are colloquially known within the asteroseismic community as Gemma, Scully, and Boogie, respectively. We denote the final star by its bright star designation, HR 7322. The inputs used for each star are summarized in Table \ref{Table:NetworkInputs}.

\begin{figure}
    \centering
	\includegraphics[width=1\columnwidth]{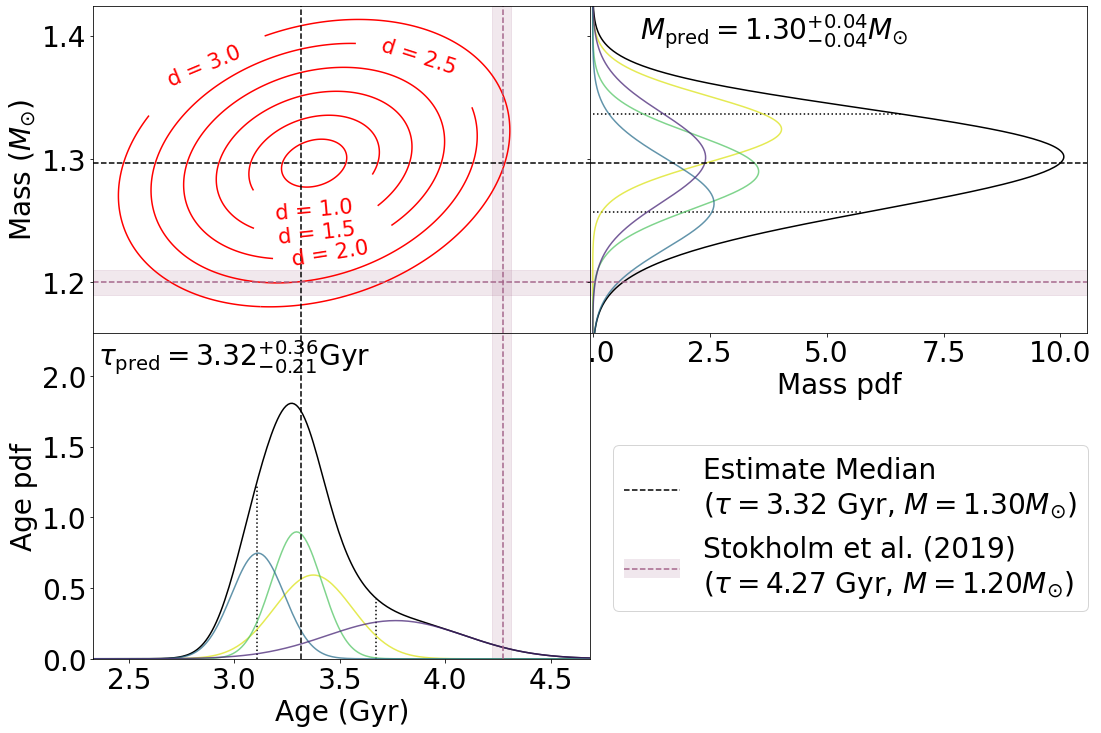}
    \caption{The estimated age and mass distribution for KIC 10005473 (HR 7322). The probability density (black) in the bottom and right panels are the network's estimates for age and mass, respectively. The black dotted lines represent the 16th and 84th percentile values. Each probability density is a superposition of up to 16 Gaussians; here only the 4 highest-weighted Gaussians are shown, with lighter colours indicating higher weights. The dashed black lines correspond to the median of the age/mass distributions. Literature values of age and mass (including uncertainties) are shaded in purple.
    The center panel shows the joint age-mass distribution, where the red contours are lines of constant Mahalanobis distance.\protect\footnotemark }
    \label{fig:contour_amalie}
\end{figure}

A comparison of our estimates with previous results from grid-based modelling is shown in Table \ref{table:SummaryGridBasedModelling}. Our estimates for $\tau$, $M$, and $R$ agree with previously modelled results for Gemma, Boogie, and Scully. The corresponding estimates for HR 7322, however, are discrepant by more than 2$\sigma$. An examination of our estimated age and mass distributions for HR 7322 in Figure \ref{fig:contour_amalie} shows that the \citet{Stokholm_2019} measurements are above the 98th percentile of our age estimate and below the 3rd percentile of our mass estimate. This indicates that the \citet{Stokholm_2019} solution is much less likely compared to a solution that is $\sim$1~Gyr younger and $\sim$0.1 $M_{\odot}$ more massive.
We note that our estimates are in excellent agreement with other mass and radius measurements reported by \citet{Stokholm_2019} for HR 7322, which are  $M=1.35\pm0.07M_{\odot}$ and $R=2.04\pm0.04R_{\odot}$ from the asteroseismic scaling relations, and the value of $R=2.00\pm0.03R_{\odot}$ from interferometry.

\subsection{Estimate Self-consistency}\label{self-consistency}
\begin{figure}
    \centering
	\includegraphics[width=1.\columnwidth]{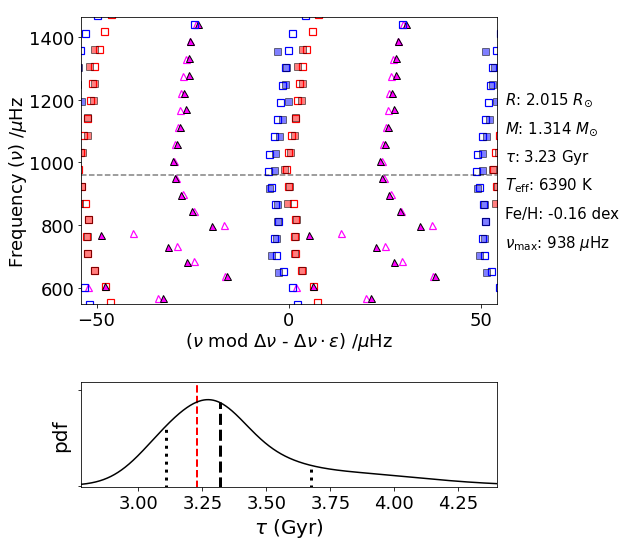}
	\vspace{-7pt}
    \caption{(Top) Model generated using initial parameters that are sampled from the estimated distribution for KIC 10005473 (HR 7322).   Model frequencies are represented by open symbols, while filled symbols represent observed frequencies. The initial parameters used to generate this model are tabulated in Table \ref{table:model_parameters}.
    (Bottom) The model's age of $\tau=3.23$~Gyr (thin red line) is located near the peak of the estimated age distribution. The thick dashed line is the distribution median and the dotted lines correspond to the 16th and 84th percentile values.}
    \label{fig:amalie_rmax}
\end{figure}

Our trained network is a deterministic function that provides estimates of stellar properties when given a set of input observables. While it is encouraging that our results in Table \ref{table:SummaryGridBasedModelling} agree well with most from grid-based modelling, it does not necessarily indicate that the estimated stellar properties are self-consistent. Machine learning algorithms only learn data-driven relations from a grid of models and do not know about the physical laws governing stellar evolution. To test for self-consistency, we identify whether models using our estimates as initial parameters can match the observed properties of the stars analyzed in this study. First, we generate a model using initial parameters ($Y_0, Z_0, \alpha_{\mathrm{MLT}}, \alpha_{\mathrm{over}}, \alpha_{\mathrm{under}}, D$) that we sample from the network's output distribution. This initial model typically has avoided crossing frequencies close to the observed avoided crossings of the star.
To improve the match between model and observation, we generate new models with the same initial parameters but with $M$ and $R$ simultaneously varied in steps of $0.1\sigma_{M}$ and $0.1\sigma_{R}$, respectively. The simultaneous variation of $M$ and $R$ preserves the root mean density, $\rho^{1/2} = \sqrt{M/R^3}$, and thus the $\Delta\nu$ of the initial estimate. Each model generated has their mode frequencies corrected for the surface term offset using Equation \ref{cubic_equation}. Using our simple search method, we identify the best-matching model by finding the model with the lowest $\chi^2$ score. The $\chi^2$ score is a measure of the goodness of fit of each model's frequencies and spectroscopic properties ($\vec{x}_{\mathrm{mod}}$) with respect to the stellar observables ($\vec{x}_{\mathrm{obs}}$) is evaluated by computing $\chi^2= (\vec{x}_{\mathrm{obs}} - \vec{x}_{\mathrm{mod}})^2/\sigma_{\mathrm{obs}}^2$, where $\sigma_{\mathrm{obs}}$ are observational uncertainties. In Figure \ref{fig:amalie_rmax}, we show an example of a model generated using the network's estimates that provide a good match to the observed properties of HR 7322. Examples of models using the network's estimates for Gemma, Scully, and Boogie are shown in Appendix \ref{Appendix:MatchingModels}, which all show good agreement with the observed properties of their corresponding subgiants.

\footnotetext{The Mahalanobis distance, $d$, is a multi-dimensional generalization of the number of standard deviations that a point $\vec{x}$ is from the mean $\vec{\mu}$ of a distribution \citep{Mahalanobis_1936}. Mathematically, it is described as $d=(\vec{x}-\vec{\mu})^TS^{-1}(\vec{x}-\vec{\mu})$, where $S$ is the covariance matrix of the distribution. In Figure \ref{fig:contour_amalie}, $d$ is used to visualize the range of values that $\vec{x} = (\tau, M)$ can have when sampling from the joint age-mass distribution.}

\subsection{Fundamental Parameter Estimation: Subgiant Ensemble}\label{Section:Ensemble}

We now apply our method on a sample of 30 oscillating \textit{Kepler} subgiant stars that were seismically analyzed by \citet{YaguangLi_2020}. Using these extracted oscillation frequencies, \citet[hereafter T20]{TandaLi_2020} used a grid of stellar models to estimate ages for each subgiant in the sample. Because they find that changes to $Y_0$ and $\alpha_{\mathrm{MLT}}$ do not strongly influence the ages of subgiant stars, they construct a grid of models varied only in $M$ and [Fe/H]. Consequently, they adopt a solar-calibrated $\alpha_{\mathrm{MLT}}$ of 1.9 and estimated $Y_0$ using the Galactic chemical evolution law. Their formulation neglects heavy element diffusion and includes an exponential overshooting scheme at the boundaries of convective cores and hydrogen-burning shells with a fixed overshooting parameter.

\begin{figure*}
    \centering
	\includegraphics[width=2.\columnwidth]{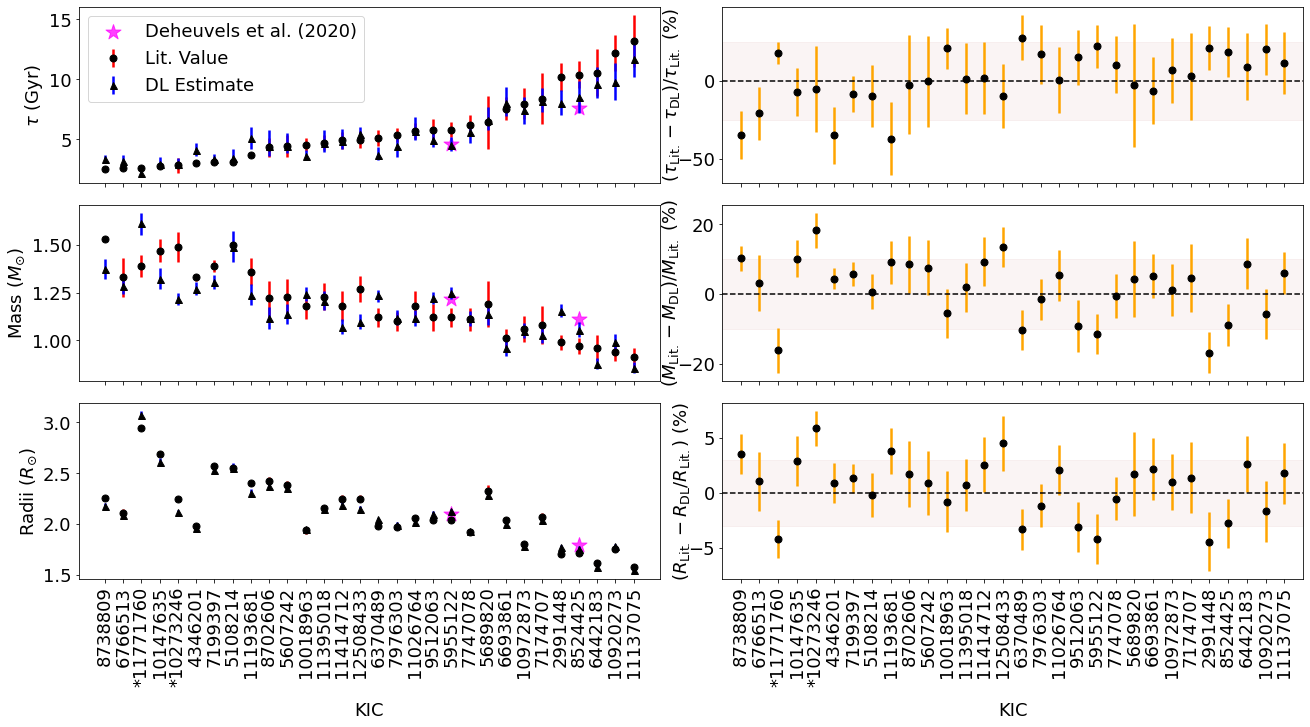}
    \caption{(Left panels) Estimates for age (top), mass (middle), and radius (bottom) values for a sample of 30 \textit{Kepler} subgiant stars in this work (blue, subscript `DL'), as compared to values inferred by \citet{TandaLi_2020} using grid-based modelling (red, subscript `Lit.'). The pink star-shaped points in the left panels correspond to model-based estimates for KIC 5955122 and KIC 8524425 by \citet{Deheuvels_2020}.
    Subgiants are sorted by increasing $\tau_{\mathrm{Lit.}}$ from left to right. The errorbars for `DL estimates' are the range of values between the 16th and 84th percentiles of the estimated distributions. (Right panels) Residuals of plots in the left. The errorbars of the residuals are the combined uncertainties from `Lit.' and `DL'. The shaded regions correspond to fractional difference intervals of 25\% for age, 10\% for mass, and 3\% for radius. Stars with IDs beginning with an asterisk (*) have fractional differences larger than 2$\sigma$ for both masses and radii.}
    \label{fig:ensemble_age_mass}
\end{figure*}

A comparison of our age, mass, and radius estimates\footnote{Estimates for all predicted parameters for this sample are tabulated in Appendix \ref{Appendix:Ensemble}.} with the grid-based modelling approach on this ensemble is shown in Figure \ref{fig:ensemble_age_mass}. Our age estimates are typically below a 25\% fractional difference to the ages from T20. Additionally, our estimates are typically below fractional differences of 10\% for masses and 3\% for radii. Stars with fractional differences in both $M$ and $R$ larger than 2$\sigma$ are marked with asterisks in Figure \ref{fig:ensemble_age_mass} and are identified as KIC 10273246 and KIC 11771760. The disagreement for KIC 11771760 is potentially due to the insufficient grid sampling from the model analyses by T20, which affected stars with $M_{\mathrm{Lit.}}>1.3$~$M_{\odot}$. We note that our fundamental parameter estimates for two subgiants in this ensemble, namely KIC 5955122 and KIC 8524425, agree with those from \citet{Deheuvels_2020} (pink star-shaped points), who had modelled such stars without convective overshooting but with microscopic diffusion enabled.

\begin{figure*}
    \centering
	\includegraphics[width=1.75\columnwidth]{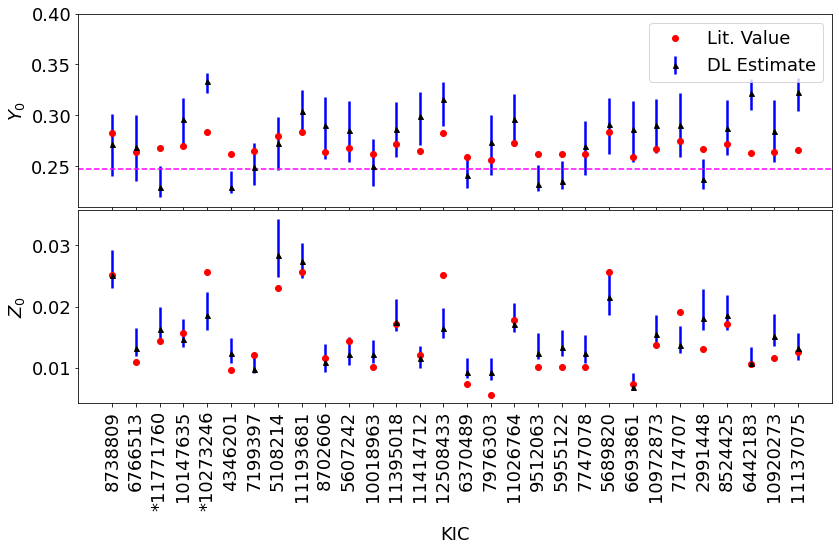}
    \caption{Comparison of initial helium abundance, $Y_0$, and the initial metal abundance, $Z_0$, for a sample of 30 \textit{Kepler} subgiant stars. The pink line in the top panel corresponds to the primordial helium abundance, $Y_p=0.2467$ \citep{Planck_2016}. Errorbars are 16th and 84th percentile values. 
    Here, stars with IDs marked with an asterisk have both mass, or radii residuals larger than their respective 2$\sigma$ values in Figure \ref{fig:ensemble_age_mass}.}
    \label{fig:ensemble_helium_alpha}
\end{figure*}

In Figure \ref{fig:ensemble_helium_alpha}, we compare our $Y_0$ and $Z_0$ estimates with the values used by T20.  
We do not find discrepancies between estimated masses, radii, or ages in Figure \ref{fig:ensemble_age_mass} to correlate strongly with differences between $Y_0$ or $Z_0$. This indicates that initial chemical abundances alone cannot account for the observed differences, and that it is likely that differences in other input physics (such as the presence/absence of microscopic diffusion and the formulation of overshooting used) play a significant role. We note, however, that the lack of correlation with $Y_0$ may be due to the insensitivity of subgiant ages to initial helium abundances, which was found by T20.
A notable observation in our estimates is the presence of 6 subgiants with an estimated median $Y_0$ marginally below the primordial helium abundance, $Y_p=0.2467$ \citep{Planck_2016}.
The occurrence of sub-primordial $Y_0$ solutions is a poorly-understood problem in fitting models of solar-like oscillators, and has been attributed to unknown systematic errors (e.g., \citealt{Mathur_2012}), or
the inadequacy of the input model physics used \citep{Bonaca_2012}. As a result, work-around methods to this problem involve artificially penalizing sub-primordial $Y_0$ solutions during a grid search (e.g., \citealt{Metcalfe_2014}) or more commonly, the use of the Galactic chemical evolution law, which effectively removes $Y_0$ as a free parameter. The prevalence of sub-primordial $Y_0$ values in our estimates may suggest that this issue cannot be solved by only having more free parameters with our current prescription of input physics in 1D stellar models. An inverse analysis, such as that which has been done for the Sun (e.g., \citealt{Basu_2016}) and main-sequence stars \citep{Bellinger_inversion_2017, Bellinger_inversion_2019} would be useful to identify missing physics from the evolutionary simulations. 


\section{Discussion}

\subsection{Additional Input Physics Parameters}\label{Section:AdditionalPhysics}
 
The purpose of including a broad range of parameters $\alpha_{\mathrm{over}}$, $D$, and $\alpha_{\mathrm{under}}$ in the grid of models in our study is to minimize the implicit assumptions of input physics when determining subgiant fundamental parameters. Our estimates are therefore expected to factor in many possible variations of input physics \textemdash{} this property is shown to some extent in Section \ref{Section:Ensemble} by the agreement of our age estimates for KIC 5955122 and KIC 8524425 with the solutions by \citet{TandaLi_2020} and \citet{Deheuvels_2020}, which both had different parameterizations of input physics. 

Additionally, by considering a range of input physics, our method has estimated that an age of $3.3$~Gyr for HR 7322 is more likely compared to its previously reported age of $4.3$~Gyr by \citet{Stokholm_2019}. In their study, \citet{Stokholm_2019} only found acceptable solutions across models of several different $\alpha_{\mathrm{MLT}}$ values and models with a fixed overshooting efficiency by allowing $Y_0$ to be less than the primordial helium abundance, $Y_p\simeq0.2467$ \citep{Planck_2016}. This problem is not encountered in our solution, where we show in Figure \ref{fig:amalie_rmax} that a model with $\tau=3.3$~Gyr and $Y_0=0.248$ shows a good match to HR 7322 by having a small amount of overshooting ($\alpha_{\mathrm{over}}\sim0.01$) with a mixing length $\alpha_{\mathrm{MLT}}\sim1.86$, which is slightly above the solar-calibrated value of 1.82.

Our estimates for $\alpha_{\mathrm{over}}$, $D$, and $\alpha_{\mathrm{under}}$ in Sections \ref{section:validation_interpretation} and \ref{Section:Boutique} show large uncertainties, indicating that these parameters cannot be easily constrained to a narrow range about a point estimate because such parameters are broadly distributed when matching models to observations. This limitation comes from the stellar models rather than from the method used in this work. In particular, the additional input physics parameters have been known to have complex effects on subgiant evolution such that a degeneracy of values can exist within a relatively narrow range of fundamental parameters when fitting subgiant models.
\citet{Deheuvels_2011} showed that there exists the possibility of having overshooting efficiencies that are either low ($\alpha_{\mathrm{over}}\apprle0.05$) or high ($\alpha_{\mathrm{over}}>0.1$), with only small differences in stellar mass. Furthermore, \citet{Deheuvels_2011} also reported that microscopic diffusion has only a subtle effect in influencing a subgiant's evolution, although it added further complexity to the interpretation of overshooting efficiencies.

Despite our estimates for the additional input physics parameters having broad distributions, we note that greater likelihoods are typically estimated for small values ($\apprle 0.1$) as can be seen from the distributions in Appendix \ref{Appendix:10d}. Indeed, the good-matching models in the analysis of self-consistency in Section \ref{self-consistency} are based on models generated with small values of the additional input physics parameters.
For core overshooting, the higher likelihood for relatively small values of $\alpha_{\mathrm{over}}$ is consistent with the analysis by \citet{Deheuvels_2011}, which showed that a moderate level of overshooting ($\alpha_{\mathrm{over}} > 0.1$) increases the proximity of a subgiant model towards the Terminal Age Main Sequence (TAMS) \textemdash{} a phase where stars are less likely to be observed. Similarly to $\alpha_{\mathrm{over}}$, $D$ and $\alpha_{\mathrm{under}}$ generally have low probability densities for relatively larger values (typically above 0.1). The circumstances under which such solutions can occur in models is beyond the scope of this paper but is planned in follow-up work.

\subsection{Interpreting Estimated Distributions}
Our deep learning method does not directly optimize the match between model and observed frequencies, as is done by conventional $\chi^2$ optimization techniques. Therefore, the mode of the estimated distribution does not necessarily provide a self-consistent model, as shown in our analysis in Figure \ref{self-consistency}. 
Because our method learns multiple realizations of input uncertainties and systematics for a given star during training, our estimates indicate credible intervals within which one realization (which is the case when measuring the properties of a subgiant star) is likely to be found. The resulting models in Figure \ref{fig:amalie_rmax} and in Appendix \ref{Appendix:MatchingModels} indeed reinforce this interpretation by showing that our estimates span regions of parameter space where good-matching models to observed subgiants can be found.

\subsection{Accuracy of Model-based Inference}
It is useful to clarify the concept of accuracy for model-derived estimates given that the analyses in this work compares our estimates with other modelling results.
There are two definitions of accuracy that are relevant to our work. The first definition measures how well the estimated stellar properties from the grid of models can approximate the most accurate determination of subgiant fundamental parameters to date, such as those from precise interferometric radii measurements. 
Because 1D stellar evolution codes have yet to fully model the physics of stellar structure and dynamics correctly, systematic differences between the interior structure of stellar models and the actual structure of the stellar interior (which can be inferred by asteroseismic inversions, e.g., \citealt{Bellinger_inversion_2017}) pose limitations to this first definition of accuracy.

The second accuracy definition relates to the ability of an optimization algorithm to search for appropriate models that fit the observed data. If an algorithm is inaccurate by this definition, it can only find poor-matching solutions even if there exists models within the grid that can closely approximate the best known measurements of an observed subgiant. By estimating distributions, our deep learning algorithm can find multiple good-matching solutions and is thus capable of being accurate by the second definition. Additionally, because our algorithm is able to efficiently search over a wide range of many free parameters, it has a greater potential in identifying a model that is accurate by the first definition compared to a method without such an ability. In contrast, fixing free parameters artificially improves the precision of the inferred subgiant properties at the cost of a potential loss in accuracy by ignoring a set of feasible solutions.

\subsection{Ensemble Analysis and Applications to TESS}
The deep learning method in this study performs well with estimating the fundamental parameters of a subgiant ensemble while only requiring very little computational time. It is therefore expected to appeal towards the inference of fundamental stellar properties over a large sample of subgiant stars. Such an inference task will be particularly valuable for characterizing stellar populations from TESS as well as those from the PLATO mission \citep{Rauer_2014} in the coming years. Except for KIC 10005473 (HR 7322), the analysis for all stars in this Section are based on \textit{Kepler} time series of observation length between 8-10 months. Thus, the network presented in this study can be readily applied to subgiant stars targeted by TESS within multiple Sectors, primarily those within the Continuous Viewing Zone.

\subsection{Further Work}
We propose in future work to extend the applicability of our method towards subgiants observed only for a month by TESS. The sparsity of detected oscillation modes, which is expected from 27-day TESS data, is a limiting factor for this version of the network. The current network's requires modes to be observed in a frequency range of at least $\pm3\Delta\nu$ around $\nu_{\mathrm{max}}$, which may not be sufficiently small for certain 1-month observations. Instead of training our network to generalize to both cases where the number of mode frequencies are sparse or plentiful, we will aim to train a network that focuses exclusively on observations where oscillation modes are sparse.

Additionally, our method motivates further exploration of convective overshoot, convective undershoot, and microscopic diffusion in subgiant stars. In particular, we will aim to establish correlations between our estimates of additional input physics parameters with a star's fundamental parameters, which will be supported by detailed stellar modelling. There is also the possibility of including additional grids that use different physical relations governing stellar evolution, which may open up the possibility of further testing scenarios such as the presence/absence of rotation, different convective overshooting schemes, or different models for convective transfer other than the mixing length theory.
Following subgiant stars, we envision in future research that fundamental parameter inference may also be attempted with deep learning algorithms for evolved red giant stars showing solar-like oscillations.

\section{Conclusions}

We have developed a deep learning algorithm that estimates the fundamental parameters of oscillating subgiant stars. By training a neural network on a grid of stellar models, our method takes as input the observed oscillation frequencies as well as spectroscopic and asteroseismic parameters, and subsequently outputs a 10D distribution comprising estimates of age, mass, radius, luminosity, the mixing length parameter, overshooting and undershooting coefficients, and the diffusion multiplier. Besides a large degree of freedom in exploring various combinations of model physics for subgiant stars, additional novelties in our approach include the use of \'echelle diagrams to represent mixed-mode patterns and the use of a mixture density network to estimate parameter distributions instead of point estimates. 

We applied our method to four oscillating subgiant stars previously modelled based on \textit{Kepler} observations of 8-10 months: KIC 11026764 (nicknamed Gemma), KIC 10920273 (nicknamed Scully), KIC 11395018 (nicknamed Boogie), and KIC 11026764 (HR 7322). Our estimates on KIC 11026764, KIC 10920273, and KIC 11395018 showed good agreement with previously modelled estimates for age, mass, and radius estimates. Our estimates for the asteroseismic benchmark subgiant star HR 7322 agree well with independent estimates from asteroseismic scaling relations and interferometry, but showed that an age of $\tau=3.3$~Gyr is more likely than the star's previously modelled estimate of $\tau=4.3$~Gyr. We determined that the values of the overshooting parameter, undershooting parameter, and the diffusion multiplier are typically difficult to constrain across subgiant stellar models because each parameter can take on a broad range of values when finding good-matching models to subgiant stars. However, smaller values of these parameters (< 0.1) are indicated to be more likely from our estimates. We showed that stellar models generated using our estimates result in good matches to the observed frequency and spectroscopic measurements for the four \textit{Kepler} subgiants we have investigated in detail. 

Finally, we estimated the fundamental parameters of a sample of 30 \textit{Kepler} subgiant stars and find good agreement with solutions obtained by traditional grid-based modelling using different prescriptions of input model physics. 
In particular, a majority of our estimates have fractional differences of below 25\% for age, below 10\% for mass, and below 3\% for radius, with 
only three stars with mass and radius discrepant above the 2$\sigma$ level. The method presented in this study brings utility to the detailed modelling of individual subgiant stars in the form of initial estimates, and can reliably determine the fundamental parameters of a large sample of subgiant stars extremely efficiently, which will be a valuable task for stellar population studies with the TESS mission.

\section*{Acknowledgements}

Funding for this Discovery mission is provided by NASA's Science Mission Directorate. We thank the entire \textit{Kepler} team without whom this investigation would not be possible. D.S. is the recipient of an Australian Research Council Future Fellowship (project number FT1400147). We acknowledge funding and support from the Stellar Astrophysics Center (SAC), which initiated this project at the Max Planck Institute for Solar System Research in G{\"o}ttingen, Germany, in cooperation with the Stellar Ages and Galactic Evolution (SAGE) group. Funding for the Stellar Astrophysics Centre is provided by The Danish National Research Foundation (Grant agreement no.: DNRF106). The research leading to the presented results has received funding from the European Research Council under the European Community's Seventh Framework Programme (FP7/2007-2013) / ERC grant agreement no. 338251 (StellarAges).
We thank Tanda Li and Yaguang Li, for interesting and fruitful discussions. We also thank J\o rgen Christensen-Dalsgaard for useful comments on the manuscript. Finally, we gratefully acknowledge the support of NVIDIA Corporation for the donation of the Titan Xp GPU used for developing the neural networks in this research.

\section*{Data Availability}
The trained neural network used to produce the results in this work as well as source code to train the network is made available at \url{https://github.com/mtyhon/deep-sub}.





\bibliographystyle{mnras}
\bibliography{bibi5} 


\appendix

\section{Network Architecture} \label{Appendix:Network_Architecture}
We detail the structure of the network in Table \ref{table:Network Structure}. The network is developed using the Pytorch version 1.1.0 deep learning library \citep{Pytorch}.
\newpage

\begin{table} 
	\centering
	\begin{threeparttable}
	\caption[ConvNet-MDN Structure]{Structure of the neural network for stellar model inference.}
	\label{table:Network Structure}
		\begin{tabular}{|p{2cm}|p{1.6cm}|p{1.6cm}|c|}
			\hline
			\centering{Component} & Layer & Weight Shape & Output Shape\\
			\hline
			\multirow{10}{2cm}{\centering Convolutional Network}
			 & conv1$^a$ & (8,5) & (128,128,8) \vspace{0.075cm}\\
			 & pool1 & - & (64,64,8) \vspace{0.075cm}\\
			 & conv2 & (16,3) & (64,64,16) \vspace{0.075cm}\\
			 & pool2 & - & (32,32,16) \vspace{0.075cm}\\
			 & conv3 & (32,3) & (32,32,32) \vspace{0.075cm}\\
			 & pool3 & - & (16,16,16) \vspace{0.075cm}\\
			 & flatten & - & 4096 \vspace{0.075cm}\\
			 & concatenate$^b$ & - & 4096$\times$9 \vspace{0.075cm}\\
			 & dense1 & (36864, 512) & 512 \vspace{0.075cm}\\
			 & dense2 & (512, 512) & 512 \vspace{0.075cm}\\
			 \hline \vspace{0.075cm}
			 \multirow{6}{2cm}{\centering Mixture Density Network} & $\mu$-dense1 & (512, 256) & 256 \vspace{0.075cm}\\
			 & $\mu$-dense (output) & (256, $10\times10$) & 512 \vspace{0.075cm}\\ 
			 & $\sigma$-dense1 & (512, 256) & 256 \vspace{0.075cm}\\
             & $\sigma$-dense (output) & (256, $10\times10$) & 512 \vspace{0.075cm}\\ 
 			 & $\pi$-dense1 & (512, 256) & 256 \vspace{0.075cm}\\
             & $\pi$-dense (output) & (256, 10) & 512 \vspace{0.075cm}\\
			 \hline
		\end{tabular}
		\begin{tablenotes}
      \small
      \item $^a$ For convolutional layers, weight shapes are in format (number of filters, receptive field size), while output shapes are in format (height, width, number of filters).
      \item $^b$ Each input observable (except the \'echelle diagram) is multiplied with a copy of the flatten layer output and concatenated with the same layer's output.
    \end{tablenotes}
  \end{threeparttable}
\end{table}

\section{Selection of Number of Gaussians}\label{Appendix:num_gaussians}
\begin{figure}
    \centering
    \includegraphics[width=0.95\columnwidth]{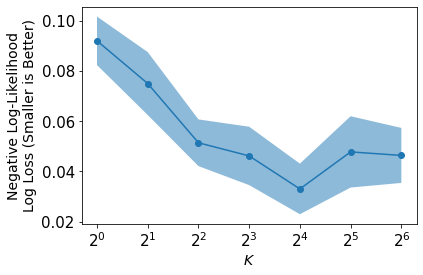}
    \caption{Variation of negative log-likelihood (Equation \ref{eq:nll}) on the validation set as a function of $K$, the number of Gaussians to model the output of the Mixture Density Network. Each value of $K$ was tested 5 times, with the average likelihood shown by the points and the envelope showing the corresponding standard deviation.}
    \label{fig:nll}
\end{figure}

We test the use of $K=1,2,4,8,16,32$, and $64$ Gaussian distributions to model the output distribution of estimates on the validation set of models in our grid as shown in Figure \ref{fig:nll}. The value of $K=16$ provides the lowest value of the negative log-likelihood, with larger values increasing the log-likelihood due to overfitting.

\section{Bootstrapping Procedure}\label{Appendix_Bootstrapping}

A summary of data generation for each training iteration is described by the following pseudo-code, with the notation (') implying perturbed quantities:
\begin{algorithmic}
\For {each training iteration}
    \For {each stellar model}
        \State Obtain mode frequencies, $\nu$
        \State Sample artificial surface term $c$ and calculate $\delta\nu_{\mathrm{surf}}$
        \State $\nu'\gets \nu - \delta\nu_{\mathrm{surf}}$
        \State Sample $\sigma_{l=0,1}$
        \State $\nu'\gets \nu' + \sigma_{l=0,1}$
        \State Calculate missing mode factor $P$
        \State $\nu'\gets P\times\nu'$
        \State Calculate $\Delta\nu$, $\epsilon$
        \State Sample $\sigma_{\nu_{\mathrm{max}}}, \sigma_{T_{\mathrm{eff}}}, \sigma_{\mathrm{[Fe/H]}}$
        \State $\nu_{\mathrm{max}}'\gets \nu_{\mathrm{max}} + \sigma_{\nu_{\mathrm{max}}}$
        \State $ T_{\mathrm{eff}}', \mathrm{[Fe/H]}'\gets T_{\mathrm{eff}} + \sigma_{T_{\mathrm{eff}}}, \mathrm{[Fe/H]} + \sigma_{\mathrm{[Fe/H]}}$
        \State Construct \'echelle diagram, $\mathcal{E}$
        \State \Return $\mathcal{E}, \nu_{\mathrm{max}}', T_{\mathrm{eff}}', \mathrm{[Fe/H]}', \sigma_{l=0,1}, \sigma_{\nu_{\mathrm{max}}}, \sigma_{T_{\mathrm{eff}}}, \sigma_{\mathrm{[Fe/H]}} $
    \EndFor
    \State Calculate network negative log-likelihood, $E$
    \State Update network weights
\EndFor
\end{algorithmic}

\section{Estimated 10D Distribution for Subgiant Stars}\label{Appendix:10d}
\begin{figure*}
    \centering
	\includegraphics[width=1.8\columnwidth]{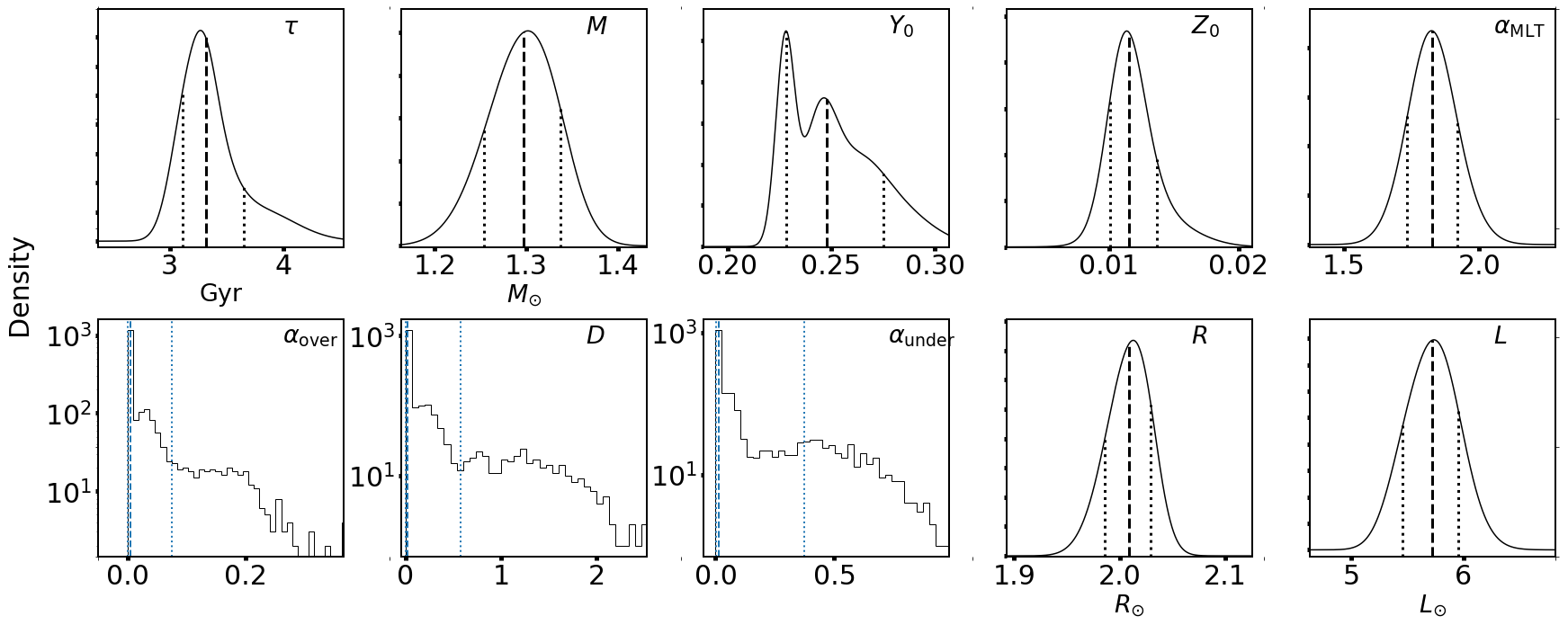}
    \caption{Probability density of each estimated parameter for KIC 10005473 (HR 7322). The black dashed lines indicate the median estimated values, with dotted black lines representing the 16th and 84th percentile values. The densities for $\alpha_{\mathrm{over}}$, $\alpha_{\mathrm{under}}$, and $D$ are plotted with logarithmic y-axes for visual clarity.}
    \label{fig:10d_amalie}
\end{figure*}

\begin{figure*}
    \centering
	\includegraphics[width=1.8\columnwidth]{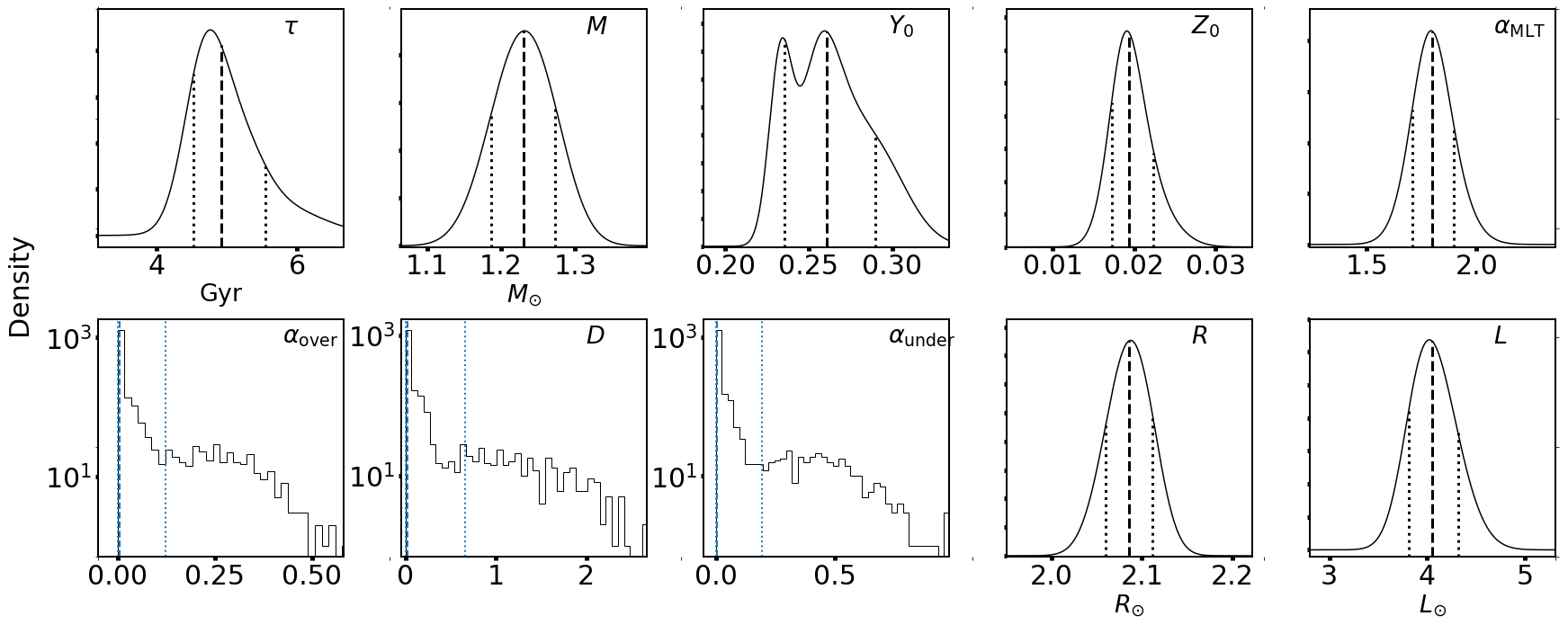}
    \caption{Same as Figure \ref{fig:10d_amalie}, but for KIC 11026764 (Gemma).}
    \label{fig:10d_gemma}
\end{figure*}

\begin{figure*}
    \centering
	\includegraphics[width=1.8\columnwidth]{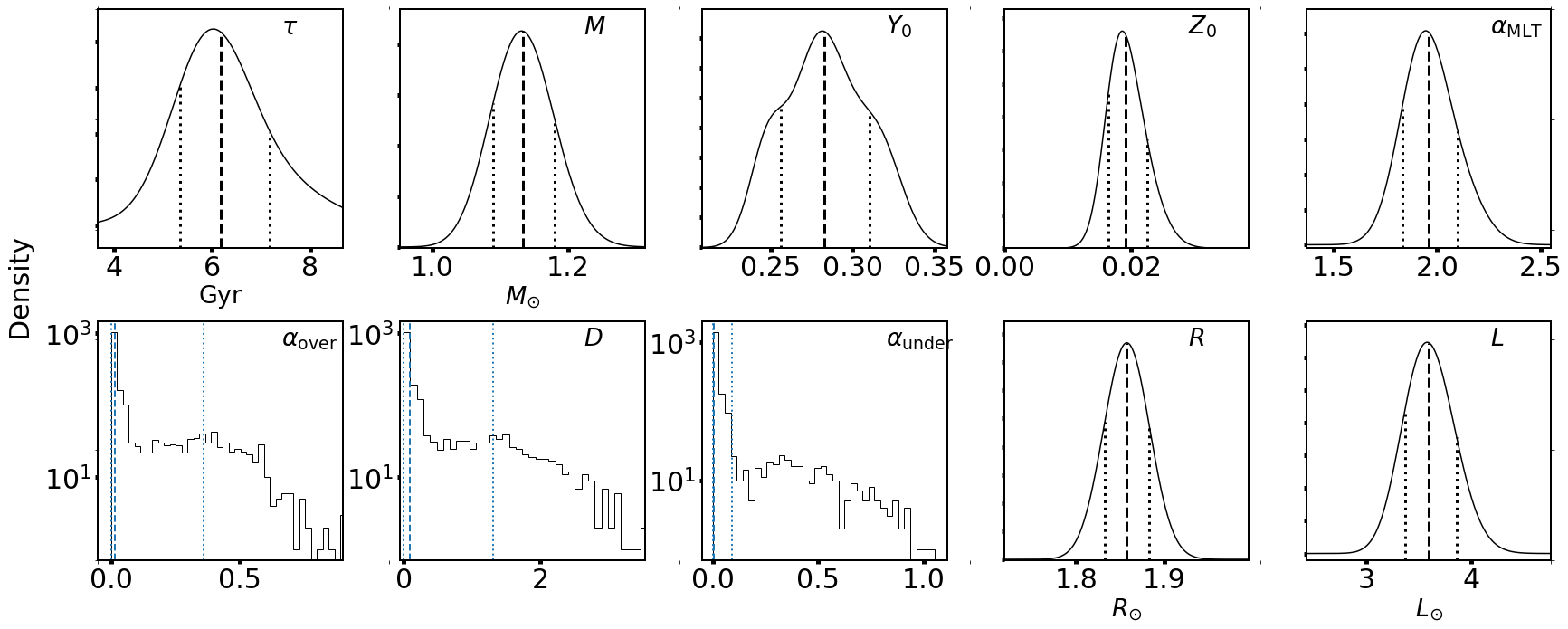}
    \caption{Same as Figure \ref{fig:10d_amalie}, but for KIC 10920273 (Scully).}
    \label{fig:10d_scully}
\end{figure*}

\begin{figure*}
    \centering
	\includegraphics[width=1.8\columnwidth]{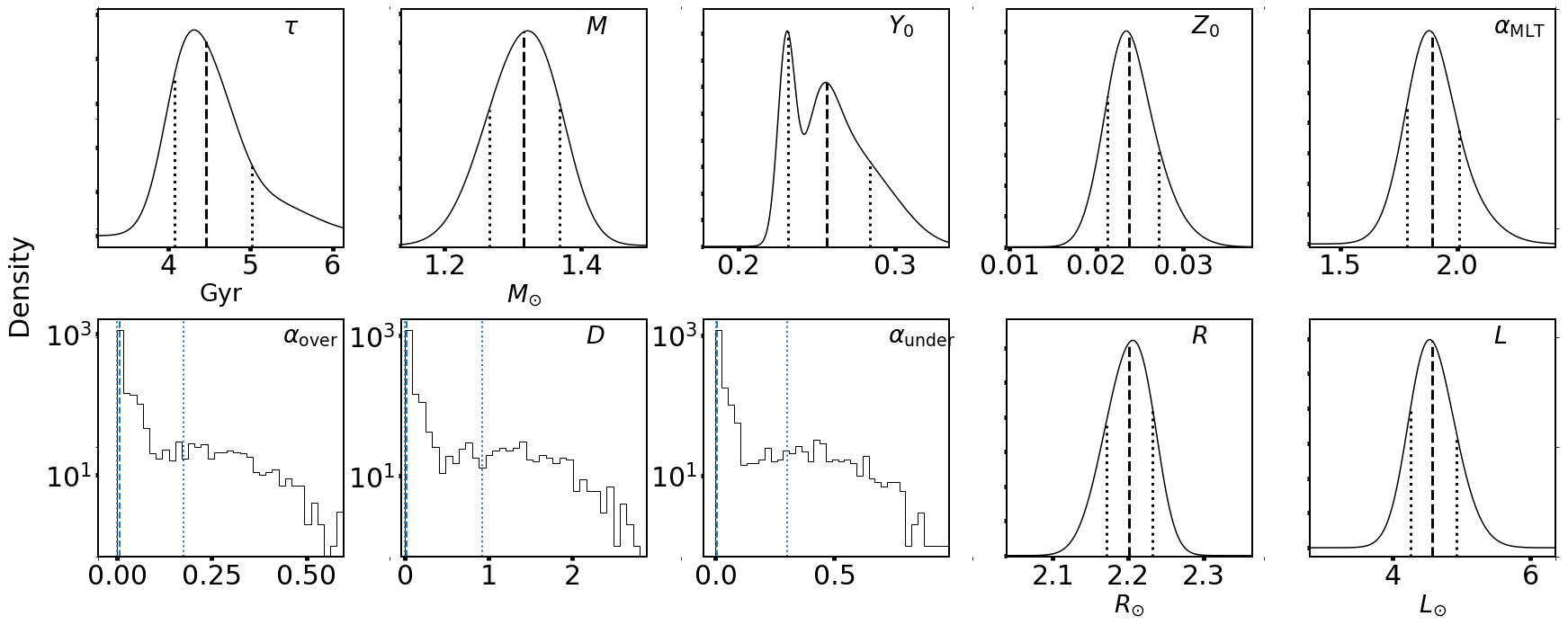}
    \caption{Same as Figure \ref{fig:10d_amalie}, but for KIC 11395018 (Boogie).}
    \label{fig:10d_boogie}
\end{figure*}

In Figures \ref{fig:10d_amalie}, \ref{fig:10d_gemma}, \ref{fig:10d_scully}, and \ref{fig:10d_boogie}, we show the full estimated probability densities for KIC 10005473 (HR 7322), KIC 11026764 (Gemma), KIC 10920273 (Scully), and KIC 11395018 (Boogie), respectively. 


\section{Matching Models using Network Estimates}\label{Appendix:MatchingModels}
\begin{figure}
    \centering
	\includegraphics[width=1.\columnwidth]{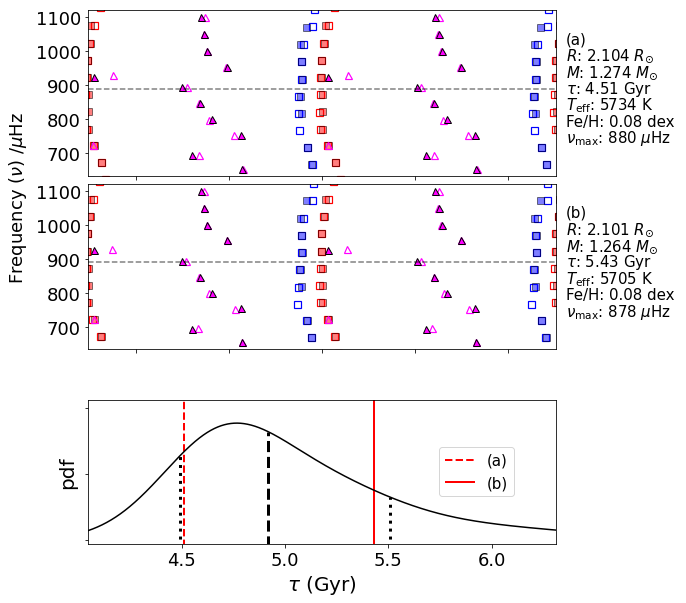}
	\vspace{-12pt}
    \caption{(Top) \'Echelle diagrams comparing model frequencies (open symbols) with observed frequencies (filled symbols) from KIC 11026764. Each panel corresponds to different initial parameters. (Bottom) A comparison of model $\tau$ with respect to the estimated age distribution.}
    \label{fig:gemma_rmax}
\end{figure}

\begin{figure}
    \centering
	\includegraphics[width=1.\columnwidth]{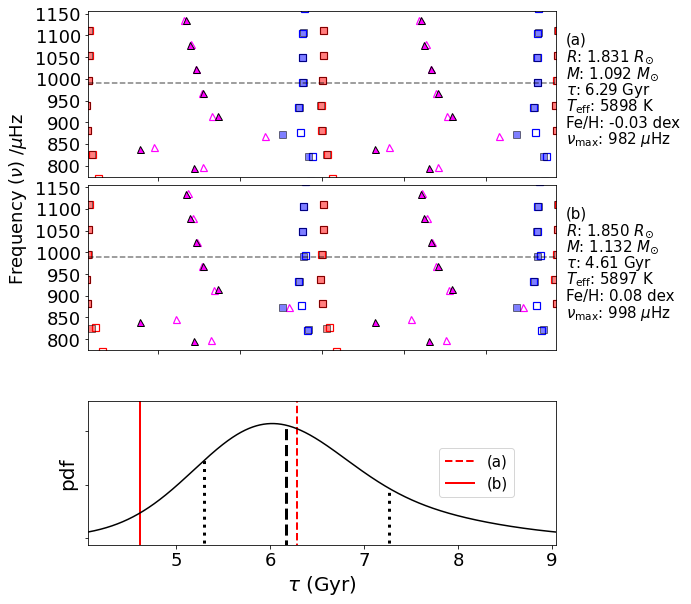}
    \caption{Same as Figure \ref{fig:gemma_rmax}, but for Scully. The initial parameters for each model is tabulated in Table.}
    \label{fig:scully_rmax}
\end{figure}

\begin{figure}
    \centering
	\includegraphics[width=1.\columnwidth]{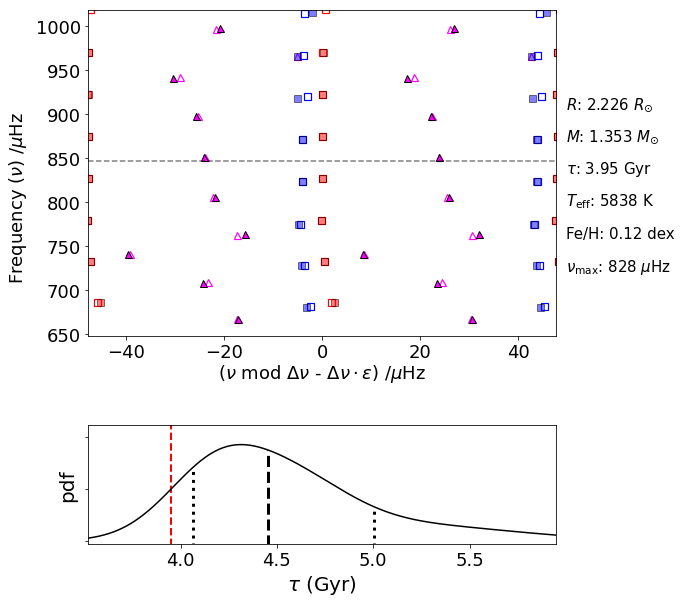}
    \caption{Same as Figure \ref{fig:gemma_rmax}, but for Boogie. The initial parameters for the model is tabulated in Table.}
    \label{fig:boogie_rmax}
\end{figure}

Figures \ref{fig:gemma_rmax}, \ref{fig:scully_rmax}, and \ref{fig:boogie_rmax} show examples of good-matching models to Gemma, Scully, and Boogie that were found using the search method described in Section \ref{self-consistency}.
Table \ref{table:model_parameters} lists the initial parameters used to generate each model.


\begin{table}
\setlength\tabcolsep{2.25pt}
	\centering
	\caption{Input parameters for the good-matching models for Gemma. The age of each model is included for reference. }
	\label{table:model_parameters}
		\begin{tabular}{|c|c|c|c|c|c|c|c|c|}
			\toprule
			Star & Age~(Gyr) & $M$~($M_{\odot}$) & $Y_0$ & $Z_0$ & $\alpha_{\mathrm{MLT}}$ & $\alpha_{\mathrm{over}}$ & $\alpha_{\mathrm{under}}$ & $D$ \\	
			\midrule
			HR 7322 & 3.23 & 1.31 & 0.248 & 0.012 & 1.86 & 0.008 & 0.014 & 0.060\\
			\hline
		    \multirow{2}{*}{Gemma} & 4.51 & 1.27 & 0.252 & 0.020 & 1.84 & 0.011 & 0.013 & 0.029\\
			& 5.43 & 1.26 & 0.253 & 0.020 & 1.81 & 0.011 & 0.010 & 0.033\\
			\hline
			\multirow{2}{*}{Scully} & 4.61 & 1.13& 0.296 & 0.019 & 2.03 & 0.017 & 0.009 & 0.025\\
			& 6.29 & 1.09 & 0.290 & 0.015 & 1.94 & 0.009 & 0.022 & 0.118\\
			\hline
			Boogie & 3.95 & 1.34 & 0.259 & 0.022 & 1.94 & 0.014 & 0.017 & 0.043\\
			\bottomrule
		\end{tabular}
\end{table}

\onecolumn{
\section{Estimated Parameters for Ensemble of 30 \textit{Kepler} Subgiant Stars}\label{Appendix:Ensemble}
In Table \ref{table:EnsembleResults}, we tabulate our full estimates on the sample of 30 \textit{Kepler} subgiant stars that were modelled by \citet{TandaLi_2020}.
}

\begin{table*} 
	\centering
	\begin{threeparttable}
	\caption{Estimates for the ensemble of 30 oscillating \textit{Kepler} subgiant stars in this study.}
	\label{table:EnsembleResults}
\begin{tabular}{|c|c|c|c|c|c|c|c|c|c|c|}
\toprule
      KIC & $\tau$ (Gyr) & $M$ ($M_{\odot}$) & $R$ ($R_{\odot}$) & $L$ ($L_{\odot}$) &  $Y_0$ & $Z_0$ &  $\alpha_{\mathrm{MLT}}$ &    $\alpha_{\mathrm{over}}$ &  $\alpha_{\mathrm{under}}$ &  $D$ \\
\midrule
  
2991448 &   $8.05^{+1.02}_{-1.04}$ &  $1.16^{+0.03}_{-0.03}$ &  $1.775^{+0.018}_{-0.019}$ &  $2.92^{+0.19}_{-0.19}$ &  $0.237^{+0.019}_{-0.011}$ &  $0.018^{+0.004}_{-0.002}$ &  $1.78^{+0.13}_{-0.11}$ &  $0.028^{+0.268}_{-0.028}$ &  $0.003^{+0.208}_{-0.004}$ &  $0.146^{+1.617}_{-0.145}$\vspace{0.075cm} \\
  4346201 &   $4.09^{+0.55}_{-0.50}$ &  $1.27^{+0.03}_{-0.03}$ &  $1.962^{+0.019}_{-0.018}$ &  $4.71^{+0.29}_{-0.28}$ &  $0.230^{+0.013}_{-0.007}$ &  $0.012^{+0.002}_{-0.002}$ &  $1.69^{+0.10}_{-0.10}$ &  $0.005^{+0.074}_{-0.005}$ &  $0.022^{+0.342}_{-0.022}$ &  $0.016^{+0.836}_{-0.016}$\vspace{0.075cm} \\
  5108214 &   $3.46^{+0.71}_{-0.54}$ &  $1.49^{+0.08}_{-0.08}$ &  $2.555^{+0.039}_{-0.044}$ &  $6.61^{+0.47}_{-0.46}$ &  $0.273^{+0.024}_{-0.028}$ &  $0.028^{+0.005}_{-0.004}$ &  $1.70^{+0.12}_{-0.11}$ &  $0.195^{+0.241}_{-0.190}$ &  $0.001^{+0.070}_{-0.001}$ &  $0.761^{+1.195}_{-0.753}$\vspace{0.075cm} \\
  5607242 &   $4.46^{+1.01}_{-0.58}$ &  $1.14^{+0.05}_{-0.05}$ &  $2.358^{+0.036}_{-0.036}$ &  $4.57^{+0.29}_{-0.26}$ &  $0.286^{+0.027}_{-0.032}$ &  $0.012^{+0.002}_{-0.002}$ &  $1.84^{+0.13}_{-0.10}$ &  $0.024^{+0.467}_{-0.024}$ &  $0.003^{+0.075}_{-0.003}$ &  $0.180^{+1.477}_{-0.179}$\vspace{0.075cm} \\
  5689820 &   $6.65^{+1.02}_{-0.90}$ &  $1.14^{+0.06}_{-0.06}$ &  $2.290^{+0.041}_{-0.042}$ &  $3.58^{+0.22}_{-0.23}$ &  $0.292^{+0.024}_{-0.030}$ &  $0.022^{+0.004}_{-0.004}$ &  $2.11^{+0.19}_{-0.18}$ &  $0.158^{+0.509}_{-0.135}$ &  $0.002^{+0.306}_{-0.002}$ &  $0.018^{+1.187}_{-0.017}$\vspace{0.075cm} \\
  5955122 &   $4.54^{+0.65}_{-0.50}$ &  $1.25^{+0.03}_{-0.03}$ &  $2.125^{+0.018}_{-0.020}$ &  $4.88^{+0.29}_{-0.28}$ &  $0.235^{+0.019}_{-0.009}$ &  $0.013^{+0.002}_{-0.002}$ &  $1.85^{+0.11}_{-0.11}$ &  $0.020^{+0.171}_{-0.020}$ &  $0.012^{+0.340}_{-0.012}$ &  $0.021^{+1.337}_{-0.021}$\vspace{0.075cm} \\
  6370489 &   $3.72^{+0.60}_{-0.46}$ &  $1.24^{+0.03}_{-0.03}$ &  $2.045^{+0.017}_{-0.022}$ &  $5.53^{+0.31}_{-0.30}$ &  $0.241^{+0.019}_{-0.013}$ &  $0.009^{+0.002}_{-0.001}$ &  $2.02^{+0.13}_{-0.12}$ &  $0.017^{+0.172}_{-0.017}$ &  $0.019^{+0.334}_{-0.019}$ &  $0.019^{+1.357}_{-0.019}$\vspace{0.075cm} \\
  6442183 &   $9.62^{+1.36}_{-1.18}$ &  $0.88^{+0.03}_{-0.03}$ &  $1.577^{+0.014}_{-0.015}$ &  $2.43^{+0.15}_{-0.19}$ &  $0.322^{+0.013}_{-0.016}$ &  $0.011^{+0.002}_{-0.001}$ &  $1.70^{+0.13}_{-0.13}$ &  $0.076^{+0.408}_{-0.075}$ &  $0.001^{+0.055}_{-0.002}$ &  $0.227^{+1.237}_{-0.226}$\vspace{0.075cm} \\
  6693861 &   $8.06^{+1.24}_{-1.09}$ &  $0.96^{+0.04}_{-0.04}$ &  $1.996^{+0.030}_{-0.031}$ &  $3.52^{+0.22}_{-0.21}$ &  $0.287^{+0.027}_{-0.033}$ &  $0.007^{+0.002}_{-0.001}$ &  $1.69^{+0.14}_{-0.12}$ &  $0.057^{+0.451}_{-0.057}$ &  $0.004^{+0.287}_{-0.004}$ &  $0.092^{+1.305}_{-0.091}$\vspace{0.075cm} \\
  6766513 &   $3.17^{+0.46}_{-0.26}$ &  $1.29^{+0.05}_{-0.05}$ &  $2.088^{+0.026}_{-0.028}$ &  $5.94^{+0.34}_{-0.35}$ &  $0.268^{+0.030}_{-0.033}$ &  $0.013^{+0.003}_{-0.002}$ &  $1.65^{+0.11}_{-0.10}$ &  $0.008^{+0.130}_{-0.008}$ &  $0.007^{+0.239}_{-0.007}$ &  $0.098^{+0.990}_{-0.097}$\vspace{0.075cm} \\
  7174707 &   $8.18^{+1.09}_{-0.94}$ &  $1.03^{+0.05}_{-0.04}$ &  $2.041^{+0.032}_{-0.032}$ &  $3.12^{+0.19}_{-0.17}$ &  $0.290^{+0.029}_{-0.033}$ &  $0.014^{+0.002}_{-0.002}$ &  $1.77^{+0.14}_{-0.11}$ &  $0.008^{+0.326}_{-0.008}$ &  $0.003^{+0.057}_{-0.003}$ &  $0.122^{+1.192}_{-0.121}$\vspace{0.075cm} \\
  7199397 &   $3.35^{+0.38}_{-0.22}$ &  $1.31^{+0.03}_{-0.04}$ &  $2.535^{+0.026}_{-0.029}$ &  $6.97^{+0.40}_{-0.37}$ &  $0.249^{+0.022}_{-0.018}$ &  $0.010^{+0.002}_{-0.001}$ &  $1.67^{+0.09}_{-0.10}$ &  $0.007^{+0.121}_{-0.007}$ &  $0.007^{+0.211}_{-0.006}$ &  $0.028^{+0.783}_{-0.027}$\vspace{0.075cm} \\
  7747078 &   $5.58^{+0.88}_{-0.88}$ &  $1.12^{+0.04}_{-0.04}$ &  $1.930^{+0.022}_{-0.022}$ &  $4.10^{+0.22}_{-0.22}$ &  $0.269^{+0.024}_{-0.029}$ &  $0.012^{+0.002}_{-0.002}$ &  $1.82^{+0.12}_{-0.11}$ &  $0.043^{+0.278}_{-0.043}$ &  $0.003^{+0.220}_{-0.004}$ &  $0.183^{+1.400}_{-0.182}$\vspace{0.075cm} \\
  7976303 &   $4.46^{+0.83}_{-0.88}$ &  $1.12^{+0.04}_{-0.04}$ &  $1.993^{+0.022}_{-0.024}$ &  $4.95^{+0.26}_{-0.26}$ &  $0.273^{+0.026}_{-0.032}$ &  $0.009^{+0.002}_{-0.002}$ &  $1.83^{+0.12}_{-0.12}$ &  $0.045^{+0.272}_{-0.045}$ &  $0.003^{+0.123}_{-0.004}$ &  $0.444^{+1.360}_{-0.443}$\vspace{0.075cm} \\
  8524425 &   $8.49^{+1.31}_{-1.25}$ &  $1.06^{+0.04}_{-0.04}$ &  $1.757^{+0.021}_{-0.022}$ &  $2.69^{+0.19}_{-0.18}$ &  $0.288^{+0.026}_{-0.028}$ &  $0.019^{+0.003}_{-0.003}$ &  $1.73^{+0.12}_{-0.11}$ &  $0.050^{+0.419}_{-0.050}$ &  $0.001^{+0.108}_{-0.002}$ &  $0.238^{+1.388}_{-0.237}$\vspace{0.075cm} \\
  8702606 &   $4.44^{+1.26}_{-0.77}$ &  $1.12^{+0.06}_{-0.05}$ &  $2.379^{+0.041}_{-0.042}$ &  $4.87^{+0.32}_{-0.30}$ &  $0.290^{+0.026}_{-0.033}$ &  $0.011^{+0.002}_{-0.002}$ &  $1.92^{+0.16}_{-0.14}$ &  $0.111^{+0.508}_{-0.110}$ &  $0.001^{+0.091}_{-0.001}$ &  $0.357^{+1.629}_{-0.355}$\vspace{0.075cm} \\
  8738809 &   $3.38^{+0.31}_{-0.25}$ &  $1.37^{+0.05}_{-0.05}$ &  $2.180^{+0.029}_{-0.030}$ &  $5.48^{+0.31}_{-0.32}$ &  $0.272^{+0.029}_{-0.032}$ &  $0.025^{+0.003}_{-0.002}$ &  $1.63^{+0.09}_{-0.09}$ &  $0.009^{+0.060}_{-0.009}$ &  $0.010^{+0.389}_{-0.009}$ &  $0.014^{+0.426}_{-0.014}$\vspace{0.075cm} \\
  9512063 &   $4.94^{+0.70}_{-0.62}$ &  $1.22^{+0.03}_{-0.03}$ &  $2.104^{+0.020}_{-0.020}$ &  $4.65^{+0.28}_{-0.27}$ &  $0.232^{+0.017}_{-0.008}$ &  $0.012^{+0.003}_{-0.001}$ &  $1.86^{+0.12}_{-0.11}$ &  $0.018^{+0.162}_{-0.018}$ &  $0.014^{+0.337}_{-0.014}$ &  $0.019^{+1.251}_{-0.019}$\vspace{0.075cm} \\
 10018963 &   $3.61^{+0.46}_{-0.34}$ &  $1.24^{+0.03}_{-0.04}$ &  $1.955^{+0.018}_{-0.021}$ &  $4.96^{+0.29}_{-0.28}$ &  $0.250^{+0.025}_{-0.021}$ &  $0.012^{+0.002}_{-0.002}$ &  $1.65^{+0.09}_{-0.10}$ &  $0.005^{+0.069}_{-0.005}$ &  $0.009^{+0.316}_{-0.009}$ &  $0.027^{+0.845}_{-0.026}$\vspace{0.075cm} \\
 10147635 &   $2.95^{+0.57}_{-0.27}$ &  $1.32^{+0.06}_{-0.05}$ &  $2.612^{+0.038}_{-0.034}$ &  $7.45^{+0.45}_{-0.45}$ &  $0.296^{+0.021}_{-0.030}$ &  $0.015^{+0.003}_{-0.002}$ &  $1.67^{+0.10}_{-0.10}$ &  $0.049^{+0.240}_{-0.049}$ &  $0.001^{+0.075}_{-0.002}$ &  $0.248^{+1.162}_{-0.247}$\vspace{0.075cm} \\
 10273246 &   $2.99^{+0.44}_{-0.27}$ &  $1.22^{+0.03}_{-0.03}$ &  $2.118^{+0.020}_{-0.018}$ &  $6.01^{+0.52}_{-0.50}$ &  $0.334^{+0.006}_{-0.013}$ &  $0.019^{+0.003}_{-0.003}$ &  $1.96^{+0.18}_{-0.15}$ &  $0.008^{+0.174}_{-0.008}$ &  $0.004^{+0.077}_{-0.004}$ &  $0.084^{+0.573}_{-0.084}$\vspace{0.075cm} \\
 10920273 &   $9.75^{+1.50}_{-1.48}$ &  $0.99^{+0.04}_{-0.04}$ &  $1.779^{+0.025}_{-0.026}$ &  $2.48^{+0.17}_{-0.17}$ &  $0.285^{+0.028}_{-0.032}$ &  $0.015^{+0.003}_{-0.002}$ &  $1.58^{+0.12}_{-0.11}$ &  $0.046^{+0.386}_{-0.045}$ &  $0.001^{+0.186}_{-0.002}$ &  $0.293^{+1.544}_{-0.291}$\vspace{0.075cm} \\
 10972873 &   $7.45^{+1.06}_{-1.16}$ &  $1.05^{+0.04}_{-0.04}$ &  $1.782^{+0.022}_{-0.022}$ &  $3.13^{+0.20}_{-0.19}$ &  $0.290^{+0.024}_{-0.028}$ &  $0.015^{+0.003}_{-0.002}$ &  $1.74^{+0.12}_{-0.12}$ &  $0.050^{+0.373}_{-0.050}$ &  $0.000^{+0.081}_{-0.001}$ &  $0.203^{+1.263}_{-0.202}$\vspace{0.075cm} \\
 11026764 &   $5.66^{+1.12}_{-0.73}$ &  $1.12^{+0.04}_{-0.04}$ &  $2.017^{+0.025}_{-0.025}$ &  $3.73^{+0.22}_{-0.21}$ &  $0.296^{+0.023}_{-0.027}$ &  $0.017^{+0.003}_{-0.002}$ &  $1.75^{+0.10}_{-0.10}$ &  $0.055^{+0.374}_{-0.054}$ &  $0.001^{+0.112}_{-0.002}$ &  $0.188^{+1.165}_{-0.186}$\vspace{0.075cm} \\
 11137075 &  $11.73^{+1.54}_{-1.57}$ &  $0.85^{+0.03}_{-0.03}$ &  $1.552^{+0.017}_{-0.018}$ &  $2.08^{+0.15}_{-0.15}$ &  $0.323^{+0.012}_{-0.018}$ &  $0.013^{+0.002}_{-0.002}$ &  $1.72^{+0.15}_{-0.13}$ &  $0.059^{+0.443}_{-0.058}$ &  $0.001^{+0.047}_{-0.002}$ &  $0.183^{+1.235}_{-0.181}$\vspace{0.075cm} \\
 11193681 &   $5.10^{+0.90}_{-0.89}$ &  $1.24^{+0.06}_{-0.05}$ &  $2.309^{+0.031}_{-0.030}$ &  $4.81^{+0.29}_{-0.29}$ &  $0.304^{+0.019}_{-0.022}$ &  $0.027^{+0.002}_{-0.003}$ &  $1.79^{+0.11}_{-0.11}$ &  $0.487^{+0.247}_{-0.472}$ &  $0.000^{+0.022}_{-0.001}$ &  $0.591^{+0.952}_{-0.579}$\vspace{0.075cm} \\
 11395018 &   $4.68^{+1.05}_{-0.70}$ &  $1.21^{+0.05}_{-0.05}$ &  $2.144^{+0.031}_{-0.032}$ &  $4.68^{+0.39}_{-0.38}$ &  $0.287^{+0.024}_{-0.029}$ &  $0.017^{+0.003}_{-0.002}$ &  $1.98^{+0.14}_{-0.13}$ &  $0.158^{+0.333}_{-0.156}$ &  $0.001^{+0.098}_{-0.001}$ &  $0.486^{+1.244}_{-0.482}$\vspace{0.075cm} \\
 11414712 &   $4.84^{+0.96}_{-0.65}$ &  $1.07^{+0.04}_{-0.04}$ &  $2.192^{+0.030}_{-0.027}$ &  $4.36^{+0.26}_{-0.24}$ &  $0.299^{+0.023}_{-0.029}$ &  $0.012^{+0.001}_{-0.002}$ &  $1.63^{+0.10}_{-0.09}$ &  $0.010^{+0.345}_{-0.010}$ &  $0.004^{+0.088}_{-0.004}$ &  $0.114^{+1.059}_{-0.113}$\vspace{0.075cm} \\
 11771760 &   $2.18^{+0.16}_{-0.15}$ &  $1.61^{+0.05}_{-0.06}$ &  $3.073^{+0.036}_{-0.043}$ &  $9.30^{+0.60}_{-0.55}$ &  $0.230^{+0.018}_{-0.010}$ &  $0.016^{+0.003}_{-0.002}$ &  $1.74^{+0.11}_{-0.10}$ &  $0.008^{+0.069}_{-0.008}$ &  $0.006^{+0.263}_{-0.005}$ &  $0.018^{+0.299}_{-0.018}$\vspace{0.075cm} \\
 12508433 &   $5.47^{+0.53}_{-0.42}$ &  $1.10^{+0.04}_{-0.04}$ &  $2.148^{+0.030}_{-0.026}$ &  $3.65^{+0.19}_{-0.18}$ &  $0.316^{+0.015}_{-0.027}$ &  $0.016^{+0.003}_{-0.002}$ &  $1.87^{+0.10}_{-0.09}$ &  $0.005^{+0.044}_{-0.005}$ &  $0.003^{+0.036}_{-0.003}$ &  $0.080^{+0.154}_{-0.079}$\vspace{0.075cm} \\
\bottomrule
\end{tabular}
  \end{threeparttable}
\end{table*}


\bsp	
\label{lastpage}
\end{document}